\documentclass[twocolumn,preprintnumbers,amsmath,amssymb]{revtex4}

\usepackage{graphicx}% Include figure files
\usepackage{dcolumn}% Align table columns on decimal point
\usepackage{bm}% bold math

\usepackage{color}
\definecolor{lightblue}{rgb}{0.2,0.2,0.7}
\definecolor{darkblue}{rgb}{0,0.25,0.5}
\definecolor{redbrown}{rgb}{0.875,0.25,0.125}
\definecolor{darkgreen}{rgb}{0,0.5,0}

\newcommand{\bra}[1]{\ensuremath{\langle #1 \vert}}
\newcommand{\ket}[1]{\ensuremath{\vert #1  \rangle}}
\newcommand{\braket}[2]{\ensuremath{\langle  #1 \vert #2  \rangle}}

\newcommand{\leftexp}[2]{{\vphantom{#2}}^{#1}{#2}}
\renewcommand{\b}[1]{\ensuremath{\mathbf{#1}}}

\newcommand{\CSF}{\ensuremath{\text{CSF}}}

\newcommand{\T}{\ensuremath{\text{T}}}

\renewcommand{\L}{\ensuremath{\text{L}}}
\newcommand{\diag}{\ensuremath{\text{diag}}}
\newcommand{\sign}{\ensuremath{\text{sign}}}
\newcommand{\bas}{\ensuremath{\text{bas}}}

\newcommand{\lin}{\ensuremath{\text{lin}}}

\newcommand{\Psib}{\ensuremath{\overline{\Psi}}}

\newcommand{\vac}{\ensuremath{\text{vac}}}

\newcommand{\exact}{\ensuremath{\text{exact}}}

\newcommand{\up}{\ensuremath{\uparrow}}
\newcommand{\dn}{\ensuremath{\downarrow}}

\newcommand{\FN}{\ensuremath{\text{FN}}}

\begin{document}

\title{Full optimization of Jastrow-Slater wave functions with application to the first-row atoms and homonuclear diatomic molecules}

\author{Julien Toulouse$^{1,2}$}
\email{julien.toulouse@upmc.fr}
\author{C. J. Umrigar$^3$}
\email{CyrusUmrigar@cornell.edu}
\affiliation{
$^1$Laboratoire de Chimie Th\'eorique - UMR 7616, Universit\'e Pierre et Marie Curie (UPMC Univ Paris 06), 75005 Paris, France.\\
$^2$Laboratoire de Chimie Th\'eorique - UMR 7616, Centre National de la Recherche Scientifique (CNRS), 75005 Paris, France.\\
$^3$Laboratory of Atomic and Solid State Physics, Cornell University, Ithaca, New York 14853, USA.
}

\date{\today}
\begin{abstract}
We pursue the development and application of the recently-introduced linear optimization method for determining the optimal linear and nonlinear parameters
of Jastrow-Slater wave functions in a variational Monte Carlo framework. In this approach, the optimal parameters are found iteratively
by diagonalizing the Hamiltonian matrix in the space spanned by the wave function and its first-order derivatives, making use of a strong
zero-variance principle. We extend the method to optimize the exponents of the basis functions, simultaneously with all the other parameters,
namely the Jastrow, configuration state function and orbital parameters. We show that the linear optimization method can be thought of as
a so-called augmented Hessian approach, which helps explain the robustness of the method and permits us to extend it to minimize a linear
combination of the energy and the energy variance. We apply the linear optimization method to obtain the complete ground-state potential energy
curve of the C$_2$ molecule up to the dissociation limit, and discuss size consistency and broken spin-symmetry issues in quantum Monte Carlo
calculations. We perform calculations of the first-row atoms and homonuclear diatomic molecules with fully optimized Jastrow-Slater wave
functions, and we demonstrate that molecular well depths can be obtained with near chemical accuracy quite systematically at the diffusion
Monte Carlo level for these systems.
\end{abstract}

\maketitle

\section{Introduction}
\label{sec:intro}

Quantum Monte Carlo (QMC) methods (see e.g. Refs.~\onlinecite{HamLesRey-BOOK-94,NigUmr-BOOK-99,FouMitNeeRaj-RMP-01}) constitute an alternative
to standard quantum chemistry approaches for accurate calculations of the electronic structure of atoms, molecules and solids. The two most
commonly used variants, variational Monte Carlo (VMC) and diffusion Monte Carlo (DMC), use a flexible trial wave function, generally consisting
for atoms and molecules of a Jastrow factor multiplied by a short expansion in configuration state functions (CSFs), each consisting of a
linear combination of Slater determinants of orbitals expanded in a localized one-particle basis. To fully benefit from the considerable
flexibility in the form of the wave function, it is crucial to be able to efficiently optimize all the parameters in these wave functions.

In recent years, a lot of effort has been devoted to developing efficient methods for optimizing a large number of parameters
in QMC wave functions. One the most effective approaches is the linear optimization method of Refs.~\onlinecite{TouUmr-JCP-07}
and~\onlinecite{UmrTouFilSorHen-PRL-07}. This is an extension of the zero-variance generalized eigenvalue equation approach of Nightingale and
Melik-Alaverdian~\cite{NigMel-PRL-01} to arbitrary nonlinear parameters, and it permits a very efficient and robust energy minimization in a VMC
framework. This method was applied successfully to the simultaneous optimization of Jastrow, CSF and orbital parameters of Jastrow-Slater wave
functions for some all-electron atoms and molecules in Refs.~\onlinecite{TouUmr-JCP-07} and~\onlinecite{TouAssUmr-JCP-07}, and for the C$_2$ and
Si$_2$ molecules with pseudopotentials in Ref.~\onlinecite{UmrTouFilSorHen-PRL-07}. It has also been applied in Ref.~\onlinecite{BroTraLopNee-JCP-07}
for optimizing Jastrow, CSF and backflow parameters to obtain very accurate wave functions for the first-row atoms.

In this paper, we pursue the development and application of the linear optimization method. We extend the method to optimize the exponents of the
basis functions, simultaneously with all the other parameters, a capacity rarely available in standard quantum chemistry methods (see, however,
Refs.~\onlinecite{TacMorSuzIgu-IJQC-98,TacTanMor-IJQC-99,TacOsa-JCP-00}). This uses a slight generalization of the parametrization
employed for Jastrow-Slater wave functions to allow nonorthogonal orbitals. Also, we show that the linear optimization method can be thought of
as a so-called augmented Hessian approach. This allows us to clearly establish the connection between the linear optimization method and
a stabilized Newton optimization method, helping to explain the robustness of the approach. Moreover, this formulation permits us to extend
the method to minimize a linear combination of the energy and the energy variance, as done in Ref.~\onlinecite{UmrFil-PRL-05} with the Newton
method. We then apply the linear optimization method to obtain the complete ground-state potential energy curve of the C$_2$ molecule up to
the dissociation limit, and discuss size consistency and broken spin-symmetry issues in QMC calculations. Finally, we perform calculations of
the first-row atoms and homonuclear diatomic molecules with fully optimized Jastrow-Slater wave functions, and we demonstrate that molecular
well depths can be obtained with near chemical accuracy quite systematically at the DMC level for these systems.

When not explicitly indicated, Hartree atomic units are assumed throughout this work.

\section{Wave function optimization}

\subsection{Wave function parametrization}

To optimize a large number of parameters of a wave function, it is important to use a convenient and efficient parametrization that eliminates
redundancies. We use an $N$-electron Jastrow-Slater wave function parametrized as (see Ref.~\onlinecite{TouUmr-JCP-07})
\begin{eqnarray}
\ket{\Psi(\b{p})} = \hat{J}(\bm{\alpha}) e^{\hat{\kappa}(\bm{\kappa},\,\bm{\zeta})} \sum_{I=1}^{N_\CSF} c_I \ket{C_I (\bm{\zeta})},
\label{Psip}
\end{eqnarray}
where $\hat{J}(\bm{\alpha})$ is a Jastrow operator, $e^{\hat{\kappa}(\bm{\kappa},\,\bm{\zeta})}$ is an orbital transformation operator,
and $\ket{C_{I}(\bm{\zeta})}$ are configuration state functions (CSFs). The parameters to be optimized, collectively designated by
$\b{p}=(\bm{\alpha},\b{c},\bm{\kappa},\bm{\zeta})$, are the Jastrow parameters $\bm{\alpha}$, the CSF parameters $\b{c}$, the orbital rotation
parameters $\bm{\kappa}$ and the basis exponent parameters $\bm{\zeta}$.

We use a flexible Jastrow factor consisting of the exponential of the sum of electron-nucleus, electron-electron and electron-electron-nucleus terms,
written as systematic polynomial and Pad\'e expansions~\cite{Umr-UNP-XX} (see also Refs.~\onlinecite{FilUmr-JCP-96,GucSanUmrJai-PRB-05}). Each
CSF is a short linear combination of products of spin-up and spin-down Slater determinants $\ket{D_{\b{k}}^{\uparrow}(\bm{\zeta})}$ and
$\ket{D_{\b{k}}^{\downarrow}(\bm{\zeta})}$

\begin{eqnarray}
\ket{C_{I}(\bm{\zeta})} = \sum_{\b{k}} d_{I,\b{k}} \ket{D_{\b{k}}^{\uparrow}(\bm{\zeta})} \ket{D_{\b{k}}^{\downarrow}(\bm{\zeta})},
\end{eqnarray}
where the coefficients $d_{I,\b{k}}$ are fully determined by the spatial and spin symmetries of the state considered. The $N_{\uparrow}$-electron and
$N_{\downarrow}$-electron Slater determinants are generated from the set of \textit{non-necessarily orthonormal} orbitals at the current optimization
step, $\ket{D_{\b{k}}^\uparrow(\bm{\zeta})} = \hat{a}^{0\dag}_{k_1\uparrow}(\bm{\zeta}) \hat{a}^{0\dag}_{k_2\uparrow}(\bm{\zeta}) \cdots \,
\hat{a}^{0\dag}_{k_{N_\uparrow} \uparrow} (\bm{\zeta}) \ket{\vac}$ and $\ket{D_{\b{k}}^\downarrow(\bm{\zeta})} = \hat{a}^{0\dag}_{k_{N_\uparrow+1}
\downarrow} (\bm{\zeta}) \hat{a}^{0\dag}_{k_{N_\uparrow+2} \downarrow} (\bm{\zeta}) \cdots \, \hat{a}^{0\dag}_{k_N \downarrow} (\bm{\zeta})
\ket{\vac}$, where $\hat{a}^{0\dag}_{k\, \sigma} (\bm{\zeta}) $ (with $\sigma=\uparrow, \downarrow$) is the fermionic creation operator for
the orbital $\ket{\phi_{k}^0(\bm{\zeta})}$ (the superscript $0$ referring to the current optimization step) in the spin-$\sigma$
determinant, and $\ket{\vac}$ is the vacuum state of second quantization. The (occupied and virtual) orbitals are written as linear
combinations of $N_{\bas}$ basis functions $\ket{\chi_{\mu}(\zeta_\mu)}$ with current coefficients $\lambda_{k,\mu}^{0}$
\begin{eqnarray}
\ket{\phi_{k}^0(\bm{\zeta})} = \sum_{\mu=1}^{N_{\bas}} \lambda_{k,\mu}^{0} \ket{\chi_{\mu}(\zeta_\mu)}.
\end{eqnarray}
Specifically, in this work, we use Slater basis functions whose expression in position representation, using spherical coordinates
$\b{r}=(r,\theta,\phi)$ around an atom position $\b{r}_a$, is
\begin{eqnarray}
\braket{\b{r}}{\chi_{\mu}(\zeta_\mu)} = N_{n_\mu}(\zeta_\mu) \, r^{n_\mu-1} \, e^{-\zeta_\mu r} \, S_{l_\mu,m_\mu}(\theta,\phi),
\label{}
\end{eqnarray}
where $N_n(\zeta)=\sqrt{(2\zeta)^{2n+1}/(2n!)}$ is the radial normalization constant and $S_{l,m}(\theta,\phi)$ are the normalized real
spherical harmonics.

Some parameters in the Jastrow factor are fixed by imposing cusp conditions~\cite{Kat-CPAM-57} on the wave function; the other Jastrow parameters
are varied freely. Due to the arbitrariness of the overall normalization of the wave function, only $N_{\CSF}-1$ CSF coefficients need be varied,
e.g. the coefficient of the first CSF is kept fixed. The only restrictions that we impose on the exponents are the equality of the exponents
of the basis functions composing the $2l+1$ components of spherical harmonics having the same $l$ (e.g., $p_x$, $p_y$ and $p_z$) and, of course, the
equality of the exponents of symmetry-equivalent basis functions on equivalent atoms. Finally, the parametrization of the orbital coefficients
through the orbital rotation parameters $\bm{\kappa}$, to which we shall come next, allows one to conveniently eliminate the redundancies due
to the invariance properties of determinants under elementary row operations.

Because straightforward variation of the exponents $\bm{\zeta}$ of the basis functions results in orbitals being nonorthogonal, in this work we use
an orbital optimization formalism that applies to nonorthogonal orbitals. The general idea of this formalism appears also in valence
bond theory (see Refs.~\onlinecite{LenBal-JCP-83,LenVerPul-MP-91,Coo-INC-02}) and is a direct generalization of the formalism for optimizing
orthonormal orbitals used in standard multiconfiguration self-consistent field (MCSCF) theory (see, e.g., Ref.~\onlinecite{HelJorOls-BOOK-02}),
and, in a QMC context in Ref.~\onlinecite{TouUmr-JCP-07}. At each optimization step, the orbitals are transformed using the operator
$e^{\hat{\kappa}(\bm{\kappa},\,\bm{\zeta})}$ where $\hat{\kappa}(\bm{\kappa},\bm{\zeta})$ is the total real singlet orbital excitation operator
\begin{eqnarray}
\hat{\kappa}(\bm{\kappa},\bm{\zeta}) = \sum_{k l} \kappa_{kl} \,  \hat{E}_{kl} (\bm{\zeta}),
\label{kappa}
\end{eqnarray}
where the parameters $\kappa_{kl}$ are nonzero only for nonredundant orbital pairs (see below) and $\hat{E}_{kl}(\bm{\zeta})$ is the singlet
excitation operator from orbital $l$ to orbital $k$
\begin{eqnarray}
\hat{E}_{kl}(\bm{\zeta})=\hat{a}_{k \uparrow}^{0\dag}(\bm{\zeta}) \hat{b}_{l \uparrow}^0(\bm{\zeta}) + \hat{a}_{k \downarrow}^{0\dag}(\bm{\zeta})
\hat{b}_{l \downarrow}^0(\bm{\zeta}),
\end{eqnarray}
where $\hat{b}_{l \sigma}^0 = \sum_q (\b{O}^{-1})_{lq} \hat{a}_{q \sigma}^0$ is the dual orbital annihilation operator at the current optimization
step written in terms of the usual annihilation operators $\hat{a}_{q \sigma}^0$ and the inverse of the overlap matrix $\b{O}$ of the orbitals
with elements $O_{lq}=\braket{\phi_l^0}{\phi_q^0}$.
The operators $\hat{a}_{k \sigma}^{\dag}$ and $\hat{b}_{l \sigma}$ satisfy the canonical
anti-commutation relations $\{\hat{a}_{k \sigma}^{0\dag},\hat{a}_{l \sigma'}^{0\dag}\} = 0$, $\{\hat{b}_{k \sigma}^0,\hat{b}_{l \sigma'}^0\} = 0$
and $\{\hat{a}_{k \sigma}^{0\dag}, \hat{b}_{l \sigma'}^0\} = \delta_{kl} \delta_{\sigma\sigma'}$. The action of the operator $\hat{E}_{kl}(\bm{\zeta})$ on a spin-$\sigma$ Slater determinant is simply to replace the $l$ spin-$\sigma$ orbital by the $k$ spin-$\sigma$ orbital. Thus, in practice, the calculation of the orbital overlap matrix $\b{O}$ is not needed. In contrast to the case of orthonormal orbitals,
the operator $\hat{\kappa}(\bm{\kappa},\bm{\zeta})$ is not anti-Hermitian and thus the operator $e^{\hat{\kappa}(\bm{\kappa},\,\bm{\zeta})}$
is not unitary. The action of this operator on a Slater determinant is seen by inserting $e^{-\hat{\kappa}(\bm{\kappa},\,\bm{\zeta})}
e^{\hat{\kappa}(\bm{\kappa},\,\bm{\zeta})}=\hat{1}$ after each orbital creation operator making up the determinant and using
$e^{\hat{\kappa}(\bm{\kappa},\,\bm{\zeta})} \ket{\vac}=\ket{\vac}$;
this leads to a new Slater determinant made with the transformed orbital
creation operators
\begin{eqnarray}
\hat{a}_{k \sigma}^{\dag}(\bm{\kappa},\,\bm{\zeta}) = e^{\hat{\kappa}(\bm{\kappa},\,\bm{\zeta})} \,\, \hat{a}_{k \sigma}^{0\dag}(\bm{\zeta})
\,\, e^{-\hat{\kappa}(\bm{\kappa},\,\bm{\zeta})},
\end{eqnarray}
and, accordingly, the corresponding transformed orbitals are
\begin{eqnarray}
\ket{\phi_k(\bm{\kappa},\,\bm{\zeta})} = e^{\hat{\kappa}(\bm{\kappa},\,\bm{\zeta})} \ket{\phi_k^{0}(\bm{\zeta})} = \sum_{l} (e^{\bm{\kappa}})_{lk}
\ket{\phi_l^{0}(\bm{\zeta})},
\label{phik}
\end{eqnarray}
where the sum is over all (occupied and virtual) orbitals, and $(e^{\bm{\kappa}})_{lk}$ are the elements of the transformation
matrix $e^{\bm{\kappa}}$ constructed as the exponential of the matrix $\bm{\kappa}$ with elements $\kappa_{kl}$.
The nonredundant orbital
excitations $l \to k$ have already been discussed in Ref.~\onlinecite{TouUmr-JCP-07}. For a single-determinant wave function, the
nonredundant excitations are: closed $\to$ open, closed $\to$ virtual and open $\to$ virtual. For a multi-configuration complete active space
(CAS)~\cite{RooTaySie-CP-80} wave function, the nonredundant excitations are: inactive $\to$ active, inactive $\to$ secondary and active $\to$
secondary. For both single-determinant and multi-determinant CAS wave functions, if $l \to k$ is an allowed excitation in Eq.~(\ref{kappa}) then
the action of the reverse excitation $k \to l$ is zero, and one can choose to impose the condition $\kappa_{lk} = - \kappa_{kl}$. In this case,
$\bm{\kappa}$ is a real anti-symmetric matrix and thus $e^{\bm{\kappa}}$ is an orthogonal matrix that simply rotates the orbitals. For a general
multiconfiguration wave function (not CAS), some active $\to$ active excitations must also be included and the action of the corresponding
reverse excitation is generally not zero. If the reverse excitation is independent, one does not have to enforce the orthogonality condition
$\kappa_{lk} = - \kappa_{kl}$, in which case the transformation of Eq.~(\ref{phik}) is no longer a rotation. Of course, in addition to these
restrictions, only excitations between orbitals of the same spatial symmetry have to be considered. We note that this orbital optimization
formalism is well suited for the use of localized orbitals, although we do not explore this possibility in this work.

We note that it is also possible to keep the orbitals exactly orthonormal when varying the basis exponents (see
Refs.~\onlinecite{TacMorSuzIgu-IJQC-98,TacTanMor-IJQC-99}). If one starts from orthonormal orbitals and uses basis functions that are
for instance symmetrically orthonormalized~\cite{Low-JCP-50,Low-AP-56}
\begin{eqnarray}
\ket{\tilde{\chi}_\mu} = \sum_{\nu=1}^{N_\bas} \, (\b{B}^{-1/2})_{\nu\mu} \, \ket{\chi_\mu},
\label{chitmu}
\end{eqnarray}
where $\b{B}$ is the overlap matrix of the basis functions with elements $B_{\nu\mu}=\braket{\chi_\nu}{\chi_\mu}$, then the orthonormality of the
orbitals is preserved during the optimization. However, this complicates the calculation of the derivatives of the wave function with respect to the
exponent parameters (in particular, one needs to calculate the derivatives of the matrix $\b{B}^{-1/2}$ as done in Ref.~\onlinecite{JorSim-JCP-83}),
and the computational effort of the optimization is significantly increased.
In practice, as discussed in Section~\ref{sec:optexp}, we have found that using orthogonalized basis functions does not significantly reduce the number of iterations needed to reach convergence.

We denote by $N_p$ the total number of parameters to be optimized. The parameters at the current optimization step are denoted by
$\b{p}^0=(\bm{\alpha}^0,\b{c}^0,\bm{\kappa}^0=\b{0},\bm{\zeta}^0)$ and the corresponding current wave function by
\begin{eqnarray}
\ket{\Psi_0} = \ket{\Psi(\b{p}^0)} = \hat{J}(\bm{\alpha}^0) \sum_{I=1}^{N_\CSF} c_I^{0} \ket{C_I(\bm{\zeta}^0)}.
\label{Psi0}
\end{eqnarray}

\subsection{First-order derivatives of the wave function}
\label{sec:derivatives}

We now give the expressions for the first-order derivatives of the wave function $\ket{\Psi(\b{p})}$ of Eq.~(\ref{Psip}) with respect to the
parameters $p_i$ at $\b{p}=\b{p}^0$
\begin{eqnarray}
\ket{\Psi_i} = \left( \frac{\partial \ket{\Psi(\b{p})}}{\partial p_i} \right)_{\b{p}=\b{p}^0},
\label{Psii}
\end{eqnarray}
which collectively designate the derivatives with respect to the Jastrow parameters
\begin{eqnarray}
\ket{\Psi_{\alpha_i}}= \frac{\partial \hat{J}(\bm{\alpha}^0)}{\partial \alpha_i} \sum_{I=1}^{N_\CSF} c_I^{0} \ket{C_I(\bm{\zeta}^0)},
\label{}
\end{eqnarray}
with respect to the CSF parameters
\begin{eqnarray}
\ket{\Psi_{c_I}} &=&  \hat{J}(\bm{\alpha}^0) \ket{C_I(\bm{\zeta}^0)},
\label{}
\end{eqnarray}
with respect to the orbital parameters
\begin{eqnarray}
\ket{\Psi_{\kappa_{kl}}} &=& \hat{J}(\bm{\alpha}^0)  \sum_{I=1}^{N_\CSF} c_I^{0} \, \hat{E}_{kl}(\bm{\zeta}^0) \, \ket{C_I(\bm{\zeta}^0)},
\label{dPsidkappa}
\end{eqnarray}
and with respect to the exponent parameters
\begin{eqnarray}
\ket{\Psi_{\zeta_\mu}} &=& \hat{J}(\bm{\alpha}^0) \sum_{I=1}^{N_\CSF} c_I^{0} \frac{\partial \ket{C_I(\bm{\zeta}^0)}}{\partial \zeta_\mu}.
\label{dPsidexp}
\end{eqnarray}
The derivatives with respect to the orbital parameters in Eq.~(\ref{dPsidkappa}) are thus simply generated by single excitations of orbitals
out of the CSFs. In the derivatives with respect to the exponent parameters in Eq.~(\ref{dPsidexp}), the orbital transformation operator
$e^{\hat{\kappa}(\bm{\kappa},\,\bm{\zeta})}$ in Eq.~\ref{Psip} does not contribute since the orbitals are transformed at each step so that we always have
$\bm{\kappa}^0=\b{0}$.

\subsection{Linear optimization method}

We use the linear optimization method of Refs.~\onlinecite{TouUmr-JCP-07} and~\onlinecite{UmrTouFilSorHen-PRL-07} to optimize all the
parameters in our wave functions. This is an extension of the zero-variance generalized eigenvalue equation approach of Nightingale and
Melik-Alaverdian~\cite{NigMel-PRL-01} to arbitrary nonlinear parameters, and it permits a very robust and efficient energy minimization in a
VMC context. We review here this approach from a somewhat different perspective and show how the method can be extended to minimize a linear
combination of the energy and the energy variance.

At each step of the optimization, the quantum-mechanical averages are computed by sampling the probability density of the current wave function
$\Psi_0(\b{R}) = \braket{\b{R}}{\Psi_0}$ in the $N$-electron position representation $\ket{\b{R}}=\ket{\b{r}_1, \b{r}_2, \cdots, \b{r}_N}$. We
will denote the statistical average of a local quantity, $f(\b{R})$, by $\left\langle f(\b{R}) \right\rangle = (1/M) \sum_{k=1}^{M} f(\b{R}_k)$
with $M$ electron configurations $\b{R}_k$.

\subsubsection{Minimization of the energy}
\label{sec:emin}

{\it Deterministic optimization method.}
The idea of the method is to iteratively:\\
(i) expand the normalized wave function
$\ket{\Psib(\b{p})}=\ket{\Psi(\b{p})}/\sqrt{\braket{\Psi(\b{p})}{\Psi(\b{p})}}$
to first order in the parameter variations $\Delta \b{p} = \b{p} - \b{p}^0$ around the current
parameters $\b{p}^0$
\begin{eqnarray}
\ket{\Psib_\lin(\b{p})} = \ket{\Psib_0} + \sum_{j=1}^{N_p} \Delta p_j \, \ket{\Psib_j},
\label{Psiblin}
\end{eqnarray}
where $\ket{\Psib_0} = \ket{\Psib(\b{p}^0)}$ is the normalized current wave function, and
\begin{eqnarray}
\ket{\Psib_j}=\frac{1}{\sqrt{\braket{\Psi_0}{\Psi_0}}} \left(\ket{\Psi_j} - \frac{\braket{\Psi_0}{\Psi_j}} {\braket{\Psi_0}{\Psi_0}} \ket{\Psi_0} \right),
\label{semiorthog}
\end{eqnarray}
are the first-order derivatives of the normalized wave function $\ket{\Psib(\b{p})}$ with respect to the parameters at $\b{p}^0$,
written in terms of the first-order derivatives $\ket{\Psi_j}$
of the unnormalized wave function $\ket{\Psi(\b{p})}$ given in Section~\ref{sec:derivatives};\\
(ii) minimize the expectation value of the
Hamiltonian $\hat{H}$ over this linear wave function with respect to the parameter variations $\Delta \b{p}$
\begin{eqnarray}
E_\lin = \min_{\Delta \b{p}} \frac{\bra{\Psib_\lin(\b{p})} \hat{H} \ket{\Psib_\lin(\b{p})}}{\braket{\Psib_\lin(\b{p})}{\Psib_\lin(\b{p})}};
\end{eqnarray}
(iii) update the current parameters as $\b{p}^0 \to \b{p}^0 +  \Delta \b{p}$.

The energy minimization step (ii) can be written in matrix notation as
\begin{eqnarray}
E_\lin =\min_{\Delta \b{p}}
\frac{\left( \begin{array}{cc}    1 & \Delta \b{p}^\T \\ \end{array} \right) \left(\begin{array}{cc} E_0 & \b{g}^\T/2\\
                              \b{g}/2 & \overline{\b{H}}\\
      \end{array} \right) \left( \begin{array}{c}    1 \\ \Delta \b{p} \\ \end{array} \right)}
{ \left( \begin{array}{cc}    1 & \Delta \b{p}^\T \\ \end{array} \right) \left( \begin{array}{cc}    1 & \b{0}^\T\\
                            \b{0} & \overline{\b{S}}\\
      \end{array} \right) \left( \begin{array}{c}    1 \\ \Delta \b{p} \\ \end{array} \right)},
\label{Elinaughess}
\end{eqnarray}
where $E_0= \bra{\Psib_0} \hat{H} \ket{\Psib_0}$ is the current energy, $\b{g}$ is the gradient of the energy with respect to the $N_p$ parameters
with components $g_i=2 \bra{\Psib_i} \hat{H} \ket{\Psib_0}$, $\overline{\b{H}}$ is the Hamiltonian matrix in the basis consisting of the $N_p$
wave function derivatives with elements $\overline{H}_{ij}=\bra{\Psib_i} \hat{H} \ket{\Psib_j}$, and $\overline{\b{S}}$ is the overlap matrix
in this basis with elements $\overline{S}_{ij}=\braket{\Psib_i}{\Psib_j}$. Clearly, the minimization of Eq.~(\ref{Elinaughess}) is equivalent
to solving the following $(N_p+1)$-dimensional generalized eigenvalue equation
\begin{eqnarray}
 \left(\begin{array}{cc} E_0 & \b{g}^\T/2\\
                              \b{g}/2 & \overline{\b{H}}\\
      \end{array} \right) \left( \begin{array}{c}    1 \\ \Delta \b{p} \\ \end{array} \right)
= E_\lin
 \left( \begin{array}{cc}    1 & \b{0}^\T\\
                            \b{0} & \overline{\b{S}}\\
      \end{array} \right) \left( \begin{array}{c}    1 \\ \Delta \b{p} \\ \end{array} \right),
\label{geneigeq}
\end{eqnarray}
for its lowest eigenvector. Each optimization step thus consists of a standard Rayleigh-Ritz approach.
When applied to the specific case of the orbital rotation parameters, this linear optimization
method is known in the quantum chemistry literature as the super configuration interaction or generalized Brillouin
theorem approach~\cite{LevBer-IJQC-68,GreCha-CPL-71,ChaGre-JCP-72,BanGre-IJQC-76}. It can also be seen as an instance
of the class of \textit{augmented Hessian} optimization methods with a particular choice of the matrix $\overline{\b{H}}$, sometimes also called rational function optimization
methods, which are based on a rational quadratic model of the function to optimize at each step rather than a quadratic one,
and which are known to be very powerful for optimizing MCSCF wave functions and molecular geometries (see, e.g.,
Refs.~\onlinecite{Len-JCP-80,Yar-CPL-81,LenLiu-JCP-81,SheShaSim-JCP-82,JenJor-JCP-84,BanAdaSimShe-JPC-85,Bak-JCC-86,KhaPanAve-IJQC-95,AngBof-IJQC-97,EckPulWer-JCC-97}).

We note that after solving Eq.~(\ref{geneigeq}), and before updating the current parameters, the parameter variations $\Delta \b{p}$ can be advantageously transformed according to Eq.~(32) of Ref.~\onlinecite{TouUmr-JCP-07} which corresponds to changing the normalization of the wave function $\ket{\Psib(\b{p})}$
\footnote{In the present work, as in Ref.~\onlinecite{TouUmr-JCP-07}, the normalization to unity of the wave function, $\ket{\Psib(\b{p})}=\ket{\Psi(\b{p})}/\sqrt{\braket{\Psi(\b{p})}{\Psi(\b{p})}}$, serves at the starting reference for the transformation of the parameter variations $\Delta \b{p}$. In contrast, in Ref.~\onlinecite{UmrTouFilSorHen-PRL-07}, the transformation was done by starting from an arbitrary normalization of the wave function. In order to be scrupulously correct, we note that in Eq.~(8) of Ref.~\onlinecite{UmrTouFilSorHen-PRL-07}, all instances of $(1-\xi)$ must be multiplied by $\sign (\braket{\Psi_\lin}{\Psi_0})$ for handling properly the improbable case where $\ket{\Psi_\lin}$ makes an obtuse angle with $\ket{\Psi_0}$. This correction does not affect the results of Ref.~\onlinecite{UmrTouFilSorHen-PRL-07}.}.

{\it Implementation in variational Monte Carlo.}
Nightingale and Melik-Alaverdian~\cite{NigMel-PRL-01} showed how to realize efficiently this energy minimization approach on a finite Monte
Carlo (MC) sample, which is not as obvious as it may seem. The procedure makes use of a strong zero-variance principle and can be described
as follows. If the current wave function and its first-order derivatives with respect to the parameters $\left\{ \ket{\Psib_0},  \ket{\Psib_1},
\cdots, \ket{\Psib_{N_p}} \right\}$ form a complete basis of the Hilbert space considered (or, less stringently, span an invariant subspace of
the Hamiltonian $\hat{H}$), then there exist optimal parameters variations $\Delta \b{p}$ so that the linear wave function of Eq.~(\ref{Psiblin})
is an exact eigenstate of $\hat{H}$, and therefore satisfies the local Schr\"odinger equation for any electron configuration $\b{R}_k$
\begin{eqnarray}
\bra{\b{R}_k} \, \, \hat{H} \left( \ket{\Psib_0} + \sum_{j=1}^{N_p} \Delta p_j \, \ket{\Psib_j} \right) =
\nonumber\\
\bra{\b{R}_k} \, \, E_\lin \left( \ket{\Psib_0} + \sum_{j=1}^{N_p} \Delta p_j \, \ket{\Psib_j}  \right),
\end{eqnarray}
where the $\Delta p_j$'s are \textit{independent} of $\b{R}_k$. Multiplying this equation by $\Psib_i(\b{R}_k)/\Psib_0(\b{R}_k)^2$ (where $i=0,1,...,N_p$) and averaging over
the $M$ points $\b{R}_k$ sampled from $\vert \Psib_0(\b{R})\vert^2$ leads to the following stochastic version of the generalized eigenvalue equation of Eq.~(\ref{geneigeq})
\begin{eqnarray}
 \left(\begin{array}{cc} \leftexp{M}{E}_0 & ^M\b{g}_R^\T/2\\
                              ^M\b{g}_L/2 & ^M\overline{\b{H}}\\
      \end{array} \right) \left( \begin{array}{c}    1 \\ \Delta \b{p} \\ \end{array} \right)
= E_\lin
 \left( \begin{array}{cc}    1 & \b{0}^\T\\
                            \b{0} & ^M\overline{\b{S}}\\
      \end{array} \right) \left( \begin{array}{c}    1 \\ \Delta \b{p} \\ \end{array} \right),
\nonumber\\
\label{geneigeqM}
\end{eqnarray}
whose lowest eigenvector solution gives the desired optimal parameter variations $\Delta \b{p}$ \textit{independently of the MC sample},
i.e. with \textit{zero variance}. In Eq.~(\ref{geneigeqM}), $\leftexp{M}{E}_0 = \left\langle E_\L(\b{R}) \right\rangle$ is the average of the
local energy $E_{\L}(\b{R}) = \bra{\b{R}} \hat{H} \ket{\Psi_0}/\Psi_0(\b{R})$ (the superscript $^M$ denoting the dependence on the MC sample),
$^M\b{g}_L$ and $^M\b{g}_R$ are two estimates of the energy gradient with components
\begin{eqnarray}
^Mg_{L,i} &=& 2 \left\langle \frac{\Psib_i (\b{R})}{\Psib_0(\b{R})} \frac{\bra{\b{R}} \hat{H} \ket{\Psib_0}}{\Psib_0(\b{R})} \right\rangle \nonumber \\
&=& 2 \Biggl[ \left\langle \frac{\Psi_i(\b{R})}{\Psi_0(\b{R})} E_\L(\b{R}) \right\rangle - \left\langle \frac{\Psi_i(\b{R})}{\Psi_0(\b{R})}
\right\rangle \left\langle E_\L(\b{R}) \right\rangle \Biggl],
\nonumber\\
\label{Hi0}
\end{eqnarray}
where $\Psib_i (\b{R})/\Psib_0(\b{R}) = \Psi_i (\b{R})/\Psi_0(\b{R}) - \left\langle \Psi_i (\b{R})/\Psi_0(\b{R}) \right\rangle$ has been used,
\begin{eqnarray}
^Mg_{R,j} &=& 2 \left\langle \frac{\bra{\b{R}} \hat{H} \ket{\Psib_j}}{\Psib_0(\b{R})} \right\rangle \nonumber \\
&=& 2 \Biggl[ \left\langle \frac{\Psi_j(\b{R})}{\Psi_0(\b{R})} E_\L(\b{R}) \right\rangle - \left\langle \frac{\Psi_j(\b{R})}{\Psi_0(\b{R})}
\right\rangle \left\langle E_\L(\b{R}) \right\rangle
\nonumber\\
&&+ \left\langle E_{\L,j}(\b{R}) \right\rangle \Biggl],
\label{H0j}
\end{eqnarray}
where $E_{\L,j}(\b{R}) = \bra{\b{R}} \hat{H} \ket{\Psi_j}/\Psi_0(\b{R}) - \left[ \Psi_j(\b{R})/\Psi_0(\b{R}) \right] E_{\L}(\b{R})$ is the
the derivative of the local energy with respect to the parameter $p_j$ (which is zero in the limit of an infinite sample),
$^M\overline{\b{H}}$ is the following \textit{nonsymmetric} estimate
of the Hamiltonian matrix
\begin{eqnarray}
^M\overline{H}_{ij} &=& \left\langle \frac{\Psib_i (\b{R})}{\Psib_0(\b{R})} \frac{\bra{\b{R}} \hat{H} \ket{\Psib_j}}{\Psib_0(\b{R})} \right\rangle
\nonumber\\
&=&\left\langle \frac{\Psi_i(\b{R})}{\Psi_0(\b{R})} \frac{\Psi_j(\b{R})}{\Psi_0(\b{R})} E_\L(\b{R}) \right\rangle
\nonumber\\
&& - \left\langle \frac{\Psi_i(\b{R})}{\Psi_0(\b{R})} \right\rangle  \left\langle \frac{\Psi_j(\b{R})}{\Psi_0(\b{R})} E_\L(\b{R}) \right\rangle
\nonumber\\
&& - \left\langle \frac{\Psi_j(\b{R})}{\Psi_0(\b{R})} \right\rangle  \left\langle \frac{\Psi_i(\b{R})}{\Psi_0(\b{R})} E_\L(\b{R}) \right\rangle
\nonumber\\
&& + \left\langle \frac{\Psi_i(\b{R})}{\Psi_0(\b{R})} \right\rangle  \left\langle \frac{\Psi_j(\b{R})}{\Psi_0(\b{R})} \right\rangle  \left\langle
E_\L(\b{R}) \right\rangle
\nonumber\\
&&+ \left\langle \frac{\Psi_i(\b{R})}{\Psi_0(\b{R})} E_{\L,j}(\b{R}) \right\rangle - \left\langle \frac{\Psi_i(\b{R})}{\Psi_0(\b{R})}
\right\rangle \left\langle E_{\L,j}(\b{R}) \right\rangle,
\nonumber\\
\label{Hij}
\end{eqnarray}
and $^M\overline{\b{S}}$ is the estimated overlap matrix
\begin{eqnarray}
^M\overline{S}_{ij}&=& \left\langle \frac{\Psib_i (\b{R})}{\Psib_0(\b{R})} \frac{\Psib_j(\b{R})}{\Psib_0(\b{R})} \right\rangle
\nonumber\\
&=&\left\langle \frac{\Psi_i(\b{R})}{\Psi_0(\b{R})} \frac{\Psi_j(\b{R})}{\Psi_0(\b{R})} \right\rangle  - \left\langle
\frac{\Psi_i(\b{R})}{\Psi_0(\b{R})} \right\rangle \left\langle \frac{\Psi_j(\b{R})}{\Psi_0(\b{R})} \right\rangle.
\nonumber\\
\label{Sij}
\end{eqnarray}

Now, in practice for nontrivial problems, the basis $\left\{ \ket{\Psib_0},  \ket{\Psib_1}, \cdots, \ket{\Psib_{N_p}} \right\}$ is never complete, and consequently
solving Eq.~(\ref{geneigeqM}) actually gives an eigenvector solution $^M\Delta \b{p}$ and associated eigenvalue $\leftexp{M}{E}_\lin$ that do
depend on the MC sample. But this solution is \textit{not} the solution that would be obtained by naively minimizing the energy of the MC sample
\begin{eqnarray}
\leftexp{M}{E}_\lin \not= \min_{\Delta \b{p}}
\frac{\left( \begin{array}{cc}    1 & \Delta \b{p}^\T \\ \end{array} \right) \left(\begin{array}{cc} \leftexp{M}{E}_0 & ^M\b{g}_R^\T/2\\
                              ^M\b{g}_L/2 & ^M\overline{\b{H}}\\
      \end{array} \right) \left( \begin{array}{c}    1 \\ \Delta \b{p} \\ \end{array} \right)}
{ \left( \begin{array}{cc}    1 & \Delta \b{p}^\T \\ \end{array} \right) \left( \begin{array}{cc}    1 & \b{0}^\T\\
                            \b{0} & ^M\overline{\b{S}}\\
      \end{array} \right) \left( \begin{array}{c}    1 \\ \Delta \b{p} \\ \end{array} \right)},
\nonumber\\
\label{ElinminM}
\end{eqnarray}
which would yield instead a generalized eigenvalue equation similar to Eq.~(\ref{geneigeqM}) but with symmetrized analogues of the energy gradients $^M\b{g}_L$, $^M\b{g}_R$ and the Hamiltonian
matrix $^M\overline{\b{H}}$. In fact, solving the generalized eigenvalue equation of Eq.~(\ref{geneigeqM}) leads to parameter variations with
statistical fluctuations about one or two orders of magnitude smaller than the parameter fluctuations
obtained using the symmetrized eigenvalue equation resulting from the minimization of
Eq.~(\ref{ElinminM}). Of course, in the limit of an infinite sample $M \to \infty$, the generalized eigenvalue equation of Eq.~(\ref{geneigeqM})
and the minimization of Eq.~(\ref{ElinminM}) become equivalent.

\subsubsection{Minimization of a linear combination of the energy and the energy variance}

In some cases, it is desirable to minimize a linear combination of the energy $E$ and the energy variance $V$: $(1-q) E + q V$ where $0 \leq
q \leq 1$. For instance, it has been shown that mixing in a small fraction of the energy variance (e.g., $q=0.05$)
in the optimization can significantly decrease the variance while sacrificing almost nothing of the energy~\cite{UmrFil-PRL-05}. Also,
there is a theoretical argument suggesting that if one wants to minimize the number of MC samples needed to obtain a fixed statistical
uncertainty on the average energy in DMC, then one should minimize in VMC a linear combination of the energy and the energy variance (usually,
the energy dominates by far in this linear combination)~\cite{Cep-JSP-86,MaDruTowNee-PRE-05}.
Indeed, in practice, the number of MC samples needed in DMC is often reduced by a few percent
by using $q=0.05$ rather then $q=0$~\cite{UmrFil-PRL-05}.
Finally, the energy variance may be more sensitive to some parameters than the energy.

The formulation of the linear optimization method as an augmented Hessian approach shows clearly how to introduce minimization of the energy
variance in the method. Suppose that, at each optimization step, we have some quadratic model of the energy variance to minimize
\begin{eqnarray}
V_\text{min} &=& \min_{\Delta \b{p}} \left\{ V_0 + \b{g}_V^\T \cdot \Delta \b{p}  + \frac{1}{2} \Delta \b{p}^\T \cdot \b{h}_V \cdot \Delta \b{p} \right\},
\label{VminN}
\end{eqnarray}
where $V_0=\bra{\Psib_0} (\hat{H} - E_0)^2 \ket{\Psib_0}$ is the energy variance of the current wave function $\ket{\Psib_0}$, $\b{g}_V$ is the
gradient of the energy variance with components $g_{V,i} = 2 \bra{\Psib_i} (\hat{H} - E_0)^2 \ket{\Psib_0}$ and $\b{h}_V$ is some approximation
to the Hessian matrix of the energy variance. Then, alternatively, one could minimize the following rational quadratic model
\begin{eqnarray}
V_\text{min} &=&\min_{\Delta \b{p}}
\frac{\left( \begin{array}{cc}    1 & \Delta \b{p}^\T \\ \end{array} \right) \left(\begin{array}{cc} V_0 & \b{g}_V^\T/2\\
                              \b{g}_V/2 & \b{h}_V/2 + V_0 \overline{\b{S}}\\
      \end{array} \right) \left( \begin{array}{c}    1 \\ \Delta \b{p} \\ \end{array} \right)}
{ \left( \begin{array}{cc}    1 & \Delta \b{p}^\T \\ \end{array} \right) \left( \begin{array}{cc}    1 & \b{0}^\T\\
                            \b{0} & \overline{\b{S}}\\
      \end{array} \right) \left( \begin{array}{c}    1 \\ \Delta \b{p} \\ \end{array} \right)},
\nonumber\\
\label{VminAH}
\end{eqnarray}
which agrees with the quadratic model in Eq.~(\ref{VminN}) up to second order in $\Delta \b{p}$, and which leads to the following generalized
eigenvalue equation
\begin{eqnarray}
 \left(\begin{array}{cc} V_0 & \b{g}_V^\T/2\\
                              \b{g}_V/2 & \b{h}_V/2 + V_0 \overline{\b{S}}\\
      \end{array} \right) \left( \begin{array}{c}    1 \\ \Delta \b{p} \\ \end{array} \right)
= V_\text{min}
 \left( \begin{array}{cc}    1 & \b{0}^\T\\
                            \b{0} & \overline{\b{S}}\\
      \end{array} \right) \left( \begin{array}{c}    1 \\ \Delta \b{p} \\ \end{array} \right).
\nonumber\\
\label{geneigeqVM}
\end{eqnarray}
On a finite MC sample, the overlap matrix $\overline{\b{S}}$ is estimated as before by the expression given in Eq.~(\ref{Sij}), the energy
variance is estimated as $\leftexp{M}{V}_0= \left\langle E_\L(\b{R})^2 \right\rangle - \left\langle E_\L(\b{R}) \right\rangle^2$, its gradient
is calculated as~\cite{UmrFil-PRL-05}
\begin{eqnarray}
^Mg_{V,i} &=& 2 \Biggl[ \left\langle E_{\L,i}(\b{R}) E_\L(\b{R}) \right\rangle - \left\langle E_{\L,i}(\b{R}) \right\rangle \left\langle
E_\L(\b{R}) \right\rangle
\nonumber\\
&& + \left\langle \frac{\Psi_i(\b{R})}{\Psi_0(\b{R})} E_\L(\b{R})^2 \right\rangle - \left\langle \frac{\Psi_i(\b{R})}{\Psi_0(\b{R})} \right\rangle
\left\langle E_\L(\b{R})^2 \right\rangle
\nonumber\\
&& - \, ^Mg_{L,i} \,  \left\langle E_\L(\b{R}) \right\rangle \Biggl],
\label{gVM}
\end{eqnarray}
and its Hessian can be approximated by the (positive-definite) Levenberg-Marquardt approximation~\cite{UmrFil-PRL-05}
\begin{eqnarray}
^Mh_{V,ij} &=& 2 \Biggl[  \left\langle E_{\L,i}(\b{R}) E_{\L,j}(\b{R}) \right\rangle  - \, ^Mg_{L,i} \,\left\langle E_{\L,j}(\b{R}) \right\rangle
\nonumber\\
&& - \, ^Mg_{L,j} \,\left\langle E_{\L,i}(\b{R}) \right\rangle + \, ^Mg_{L,i}  \, ^Mg_{L,j} \Biggl].
\label{hVLM}
\end{eqnarray}
We have also tested the use of the exact Hessian of the energy variance of the linear wave function of Eq.~(\ref{Psiblin}), but this Hessian
containing a quadricovariance term tends to be more noisy than the simpler Hessian of Eq.~(\ref{hVLM}), is not guaranteed to be positive
definite, and leads to a less efficient optimization.

It is clear that an arbitrary linear combination of the energy and the energy variance can be minimized by combining the augmented Hessian matrix
of the energy on the left-hand-side of Eq.~(\ref{geneigeqM}) with the augmented Hessian matrix of the energy variance on the left-hand-side of
Eq.~(\ref{geneigeqVM}) with the estimators of Eqs.~(\ref{gVM}) and~(\ref{hVLM}). We note however that this procedure destroys the zero-variance
principle described in the previous section which holds if only the energy is minimized.
In practice, introducing only a small fraction ($\sim 5\%$) of the
energy variance does not adversely affect the benefit gained from the zero-variance principle, and in fact in most cases
makes the optimization converge more rapidly.
We note that if we were to undertake the additional computational effort of computing $\bra{\b{R}} \hat{H}^2 \ket{\Psib_i}$ then it is possible
to formulate an optimization method that obeys the strong zero-variance principle even when optimizing the energy variance
or a linear combination of the energy and the energy variance.

In fact, following this procedure, any penalty function imposing some constraint, for which we have estimates of the gradient and of some approximation to the Hessian,
can be added to the energy and optimized with the linear method, hopefully without spoiling very much the benefit of the zero-variance
principle for the energy.

\subsubsection{Robustness of the optimization}
\label{robustness}

The linear optimization method has been found to be somewhat more robust than the Newton optimization
method of Ref.~\onlinecite{Sor-PRB-05} using an approximate Hessian, and even of that of Ref.~\onlinecite{UmrFil-PRL-05} using the exact Hessian, for optimizing QMC wave functions. On a finite MC sample, when minimizing
the energy, the zero-variance principle of Nightingale and Melik-Alaverdian~\cite{NigMel-PRL-01} is certainly a major ingredient in the robustness
of the method. But even in the limit of an infinite MC sample, augmented Hessian approaches are known in the quantum chemistry literature to be
more robust than the simple Newton method. In fact, it has been shown that augmented Hessian methods can have a greater radius of convergence
than the Newton method~\cite{SheShaSim-JCP-82}. This can be understood by rewriting the generalized eigenvalue equation~(\ref{geneigeq}) as
(see Refs.~\onlinecite{SheShaSim-JCP-82,BanAdaSimShe-JPC-85})
\begin{subequations}
\begin{equation}
\left( \b{h} - 2 \Delta E \, \overline{\b{S}} \right) \cdot \Delta \b{p} = - \b{g},
\end{equation}
\begin{equation}
2 \Delta E =  \b{g}^\T \cdot \Delta \b{p},
\label{DeltaE}
\end{equation}
\label{Newtonshifted}
\end{subequations}
where $\b{h} = 2 (\overline{\b{H}} -E_0 \overline{\b{S}})$ is a $N_p$-dimensional matrix and $\Delta E = E_\lin - E_0<0$ is the energy
stabilization obtained from going from the current wave function $\ket{\Psib_0}$ to the linear wave function $\ket{\Psib_\lin}$ (which is necessarily
negative within statistical noise). Equations~(\ref{Newtonshifted}) show that the linear optimization method is equivalent to a Newton method
with an approximate Hessian matrix $\b{h}$ which is level-shifted by the positive-definite matrix $- 2 \Delta E \, \overline{\b{S}}$, which
acts as a stabilizer. This method is thus closely related to the stabilized, approximate Newton method of Ref.~\onlinecite{Sor-PRB-05}. The
advantage of the present approach, beside the previously discussed zero-variance principle, is that the stabilization constant $- 2 \Delta E$
is automatically determined from the solution of the generalized eigenvalue equation. Clearly, from Eq.~(\ref{DeltaE}), $- 2 \Delta E$ tends
tends to be large far from the minimum, and tends to zero at convergence.

In some cases, when the initial parameters are very bad or the MC sample is not large enough, it is necessary to further stabilize the
optimization. A variety of different stabilization schemes are conceivable. In practice, we have found that adding a positive
constant $a_\diag$ to the diagonal of the Hamiltonian matrix, i.e. $\overline{\b{H}} \to \overline{\b{H}} + a_\diag \b{I}$ where $\b{I}$
is the identity matrix, works well. The value of $a_\diag$ is adjusted at each optimization step by performing three very short MC calculations using
correlated sampling with wave function parameters obtained with three trial values of $a_{\diag}$ and predicting by parabolic interpolation
the value of $a_{\diag}$ that minimizes the energy, with some bounds imposed. In addition, $a_{\diag}$ is forced to increase if the norm of
linear wave function variation $||\sum_{j=1}^{N_p} \Delta p_j \ket{\Psib_j}||$ or the norm of the parameter variations $||\Delta \b{p}||$
exceed some chosen thresholds, or if some parameters exit their allowed domain of variation (e.g., if a basis exponent becomes negative).

\section{Computational details}

We now give some details of the calculations that we have performed on the first-row atoms and homonuclear diatomic molecules in their ground states.

We start by generating a standard {\it ab initio} wave function using the quantum chemistry program
GAMESS~\cite{SchBalBoaElbGorJenKosMatNguSuWinDupMon-JCC-93}, typically a restricted Hartree-Fock (RHF) wave function or a MCSCF wave function
with a complete active space generated by distributing $n$ valence electrons in $m$ valence orbitals [CAS($n$,$m$)]. We use the CVB1 Slater
basis of Ema {\it et al.}~\cite{EmaGarRamLopFerMeiPal-JCC-03}, each Slater function being actually approximated by a fit to 14 Gaussian
functions~\cite{HehStePop-JCP-69,Ste-JCP-70,KolReiAss-JJJ-XX} in GAMESS.

This standard {\it ab initio} wave function is then multiplied by a Jastrow factor, imposing the electron-electron cusp condition, but with
essentially all other free parameters chosen to be initially zero to form our starting trial Jastrow-Slater wave function, and QMC calculations
are performed with the program CHAMP~\cite{Cha-PROG-XX} using the true Slater basis set rather than its Gaussian expansion. The Jastrow,
CSF, orbital and exponents parameters are simultaneously optimized with the linear energy minimization method in VMC, using an accelerated
Metropolis algorithm~\cite{Umr-PRL-93,Umr-INC-99}. We usually start with 10000 MC iterations for the first optimization step and then this
number is progressively increased at each step (typically by a factor $1.5$ to $4$) during the optimization until the energy is converged to $10^{-4}$ (for the lighter systems) or $10^{-3}$
(for the heavier systems) Hartree within a statistical accuracy of  $5.10^{-5}$ or $5.10^{-4}$, respectively. The optimization typically
converges in less than 10 steps. Once the trial wave function has been optimized, we perform a DMC calculation within the short-time and fixed-node (FN)
approximations (see, e.g., Refs.~\onlinecite{GriSto-JCP-71,And-JCP-75,And-JCP-76,ReyCepAldLes-JCP-82,MosSchLeeKal-JCP-82}).
We use an
imaginary time step of usually $\tau=0.01$ Hartree$^{-1}$ in an efficient DMC algorithm featuring very small time-step errors~\cite{UmrNigRun-JCP-93}.
For Ne and Ne$_2$ we computed the DMC energies at four time steps,
0.020, 0.015, 0.01, and 0.005 Hartree$^{-1}$, and performed an extrapolation to zero time step.

We found it convenient to start from the CVB1 basis exponents and from CSF and orbitals coefficients generated by GAMESS, but in fact the ability to
optimize all these parameters in QMC allows us to also start from cruder starting parameters without relying on a external quantum chemistry program.
In this case, a larger number of optimization iterations are needed to achieve convergence.

\section{Optimization of the basis function exponents}
\label{sec:optexp}

In Ref.~\onlinecite{TouUmr-JCP-07}, the optimization of the Jastrow, CSF and orbital parameters has been discussed in detail. We complete
here the discussion with the optimization of the exponent parameters.

\begin{figure}
\includegraphics[scale=0.35,angle=-90]{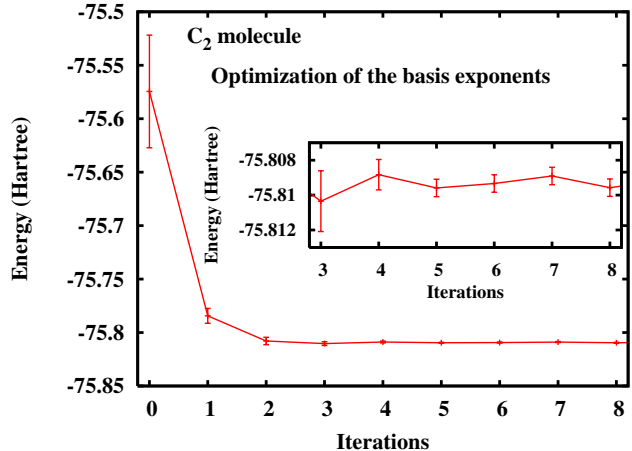}
\caption{Convergence of the VMC total energy of the all-electron C$_2$ molecule during the optimization of the 12 exponent parameters in a
wave function composed of a single Slater determinant multiplied by a Jastrow factor. Crude initial exponents have been intentionally chosen
as the integers nearest to the exponent values of the CVB1 basis~\cite{EmaGarRamLopFerMeiPal-JCC-03}. In this calculation, the number of MC sample was initially 10000 and this number was progressively increased at each iteration until a final statistical uncertainty of $0.5$ mHartree was reached.
}
\label{fig:c2_cvb1_hfj_scalek0.8_vmc_emin_e_opt_lin_norm_conv4}
\end{figure}

Figure~\ref{fig:c2_cvb1_hfj_scalek0.8_vmc_emin_e_opt_lin_norm_conv4} shows the convergence of the VMC total energy of the all-electron C$_2$
molecule during the optimization of the 12 exponent parameters in a wave function composed of a Jastrow factor multiplied by a single Slater
determinant where the Jastrow and orbital parameters have been previously optimized (with the exponents fixed at the CVB1 values).
Crude initial exponents have been intentionally chosen as
the integers nearest to the exponent values of the CVB1 basis. One sees that the linear energy minimization
method yields a fast convergence of the energy in about three iterations, typically as fast as when optimizing the other parameters. The
simultaneous optimization of the Jastrow, CSF, orbital and exponent parameters generally converges as fast as the simultaneous optimization
of only the Jastrow, CSF and orbital parameters reported in Ref.~\onlinecite{TouUmr-JCP-07}.

When optimizing the exponents without simultaneous optimization of the orbitals, we have found the optimization very stable, the introduction
of the stabilization constant $a_\diag$ often being unnecessary. However, when optimizing simultaneously the orbital and exponents parameters,
the optimization tends to be less stable because of near redundancies between these two sets of parameters, and $a_\diag$ typically increases
up to $10^{-3}-10^{-4}$ to retain stability.

Tests on a few atoms have shown that, for the optimization of exponents only, the use of the orthonormalized basis functions of Eq.~(\ref{chitmu})
tends to be a bit more stable. Typically, the overlap matrix $\overline{\b{S}}$ of the wave function derivatives have eigenvalues that span
about 7 orders of magnitude from $1$ to $10^{-7}$ for (unnormalized or normalized) nonorthogonalized functions whereas, for orthonormalized basis functions, the eigenvalues span only about 4 orders of magnitude from $1$ to $10^{-4}$.
Thus, orthogonalization of
the basis functions reduces the range of the eigenvalues, attenuating near redundancies among the exponent parameters. However, the
wave function derivatives with respect to the exponents take significantly longer to compute when using orthonormalized basis functions, and
in addition, when also optimizing the orbitals, the near redundancies between some orbital and exponents parameters make $a_\diag$ increase
anyway. Thus, we have not found it worthwhile for our purpose to use orthonormalized basis functions.

For the first-row atoms, optimizing the exponents (simultaneously with the other parameters) rather than using the exponents of the CVB1 basis
typically yields improvements of the total VMC energies between 0.1 and 1 mHartree, which is at the edge of statistical significance and
accuracy of the optimization. Thus, at the accuracy that we are concerned with, the exponents of the CVB1 basis for these atoms are nearly optimal
for Jastrow-Slater wave functions. As expected, larger improvements are obtained for the first-row diatomic molecules. The
largest improvement of the total energy is observed for the O$_2$ molecule with a gain of about $3.4$ mHartree in VMC and $1.2$ mHartree in DMC with a Jastrow-Slater single-determinant wave function.
The largest improvement of the standard deviation of the energy is obtained for the Be$_2$ dimer using a
Jastrow-Slater single-determinant wave function, with a gain of $0.25$ Hartree.
(The standard deviation of the energy is $\sigma=0.57$ Hartree with the CVB1 exponents and $\sigma=0.32$ Hartree with the reoptimized exponents
using pure energy minimization. A mixed minimization with an energy weighting of 0.95 and a variance weighting of 0.05 and reoptimized exponents
results in the same energy but a $\sigma$ of 0.32, whereas a pure variance minimization yields a variational energy
that is higher by 1 mHartree and a $\sigma$ of only 0.24.)
Properties other than the
energy might be more sensitive to the basis exponents. We note that the optimization method that we are using is designed to find local minima,
and we cannot be sure that we have found the global minimum for the form of the trial wave function considered.
In particular, optimization of the exponent parameters typically leads to multiple local minima.
We have found that by optimizing the exponents it is possible to reduce the size of the basis,
without sacrificing the energy or the energy variance, but the results in this paper were all obtained
using a basis size corresponding to the CVB1 basis. A smaller basis has also the advantage of having fewer local minima.

As noted in Ref.~\onlinecite{TacTanMor-IJQC-99}, because the virial theorem within the Born-Oppenheimer approximation at the equilibrium nuclear geometry holds if the energy is stationary with respect to scaling of the electron coordinates, optimization of the basis exponents along with optimization of the scaling factors of the interelectron coordinates in the Jastrow factor permits one to satisfy exactly the virial theorem in VMC in the limit of infinite sample size.

\section{Potential energy curve of the C$_2$ molecule}

\begin{figure*}
\includegraphics[scale=0.35,angle=-90]{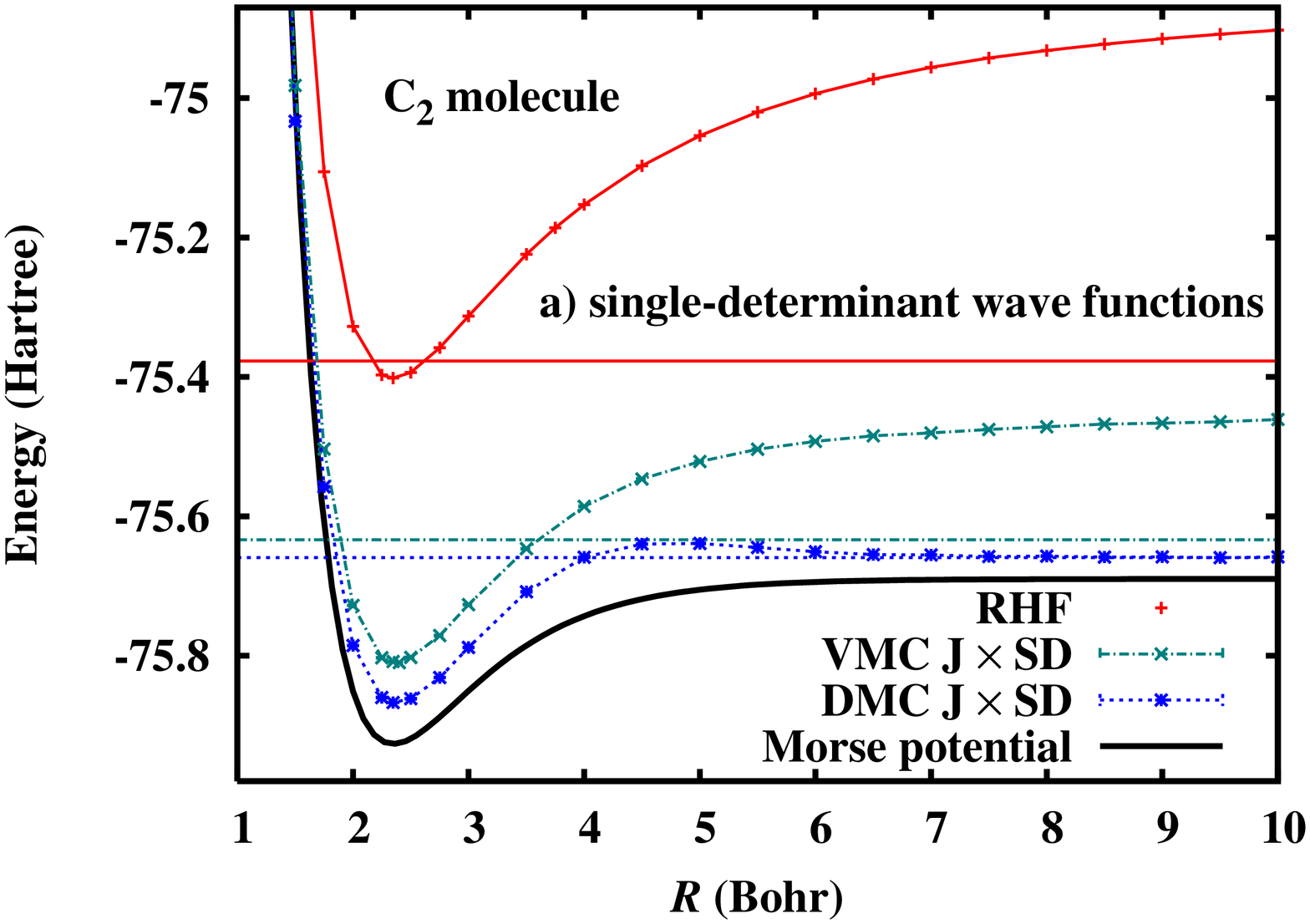}
\includegraphics[scale=0.35,angle=-90]{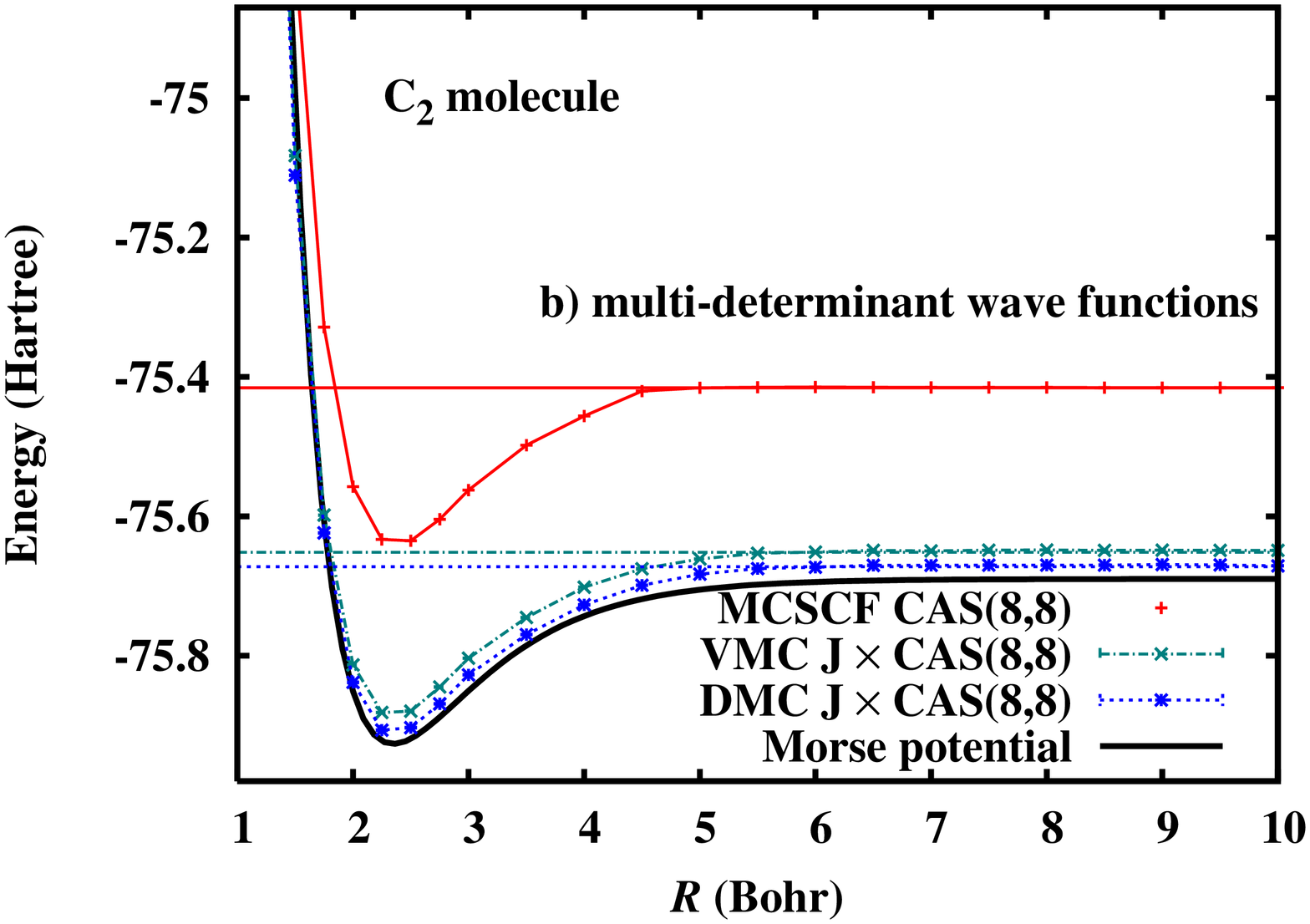}
\caption{Total energy of the all-electron C$_2$ molecule as a function of the interatomic distance $R$ calculated in (plot a) RHF, VMC
and DMC with a fully optimized Jastrow $\times$ single determinant wave function [J $\times$ SD], and (plot b) MCSCF CAS(8,8), VMC and
DMC with a fully optimized Jastrow $\times$ multi-determinant CAS(8,8) wave function [J $\times$ CAS(8,8)], using the CVB1 Slater basis
form~\cite{EmaGarRamLopFerMeiPal-JCC-03}. In each case, the horizontal line represents twice the energy of an isolated atom calculated with
the same method, and provides a check of the size consistency of the method. For comparison, a Morse potential~\cite{Mor-PR-29} using accurate
spectroscopic constants is also shown (see text).
}
\label{fig:c2_cvb1_emin_oj_evsr}
\end{figure*}

%%%%%%%%%%%%%%%%%%%%%%%%%%%%%%%%%%%%%%%%%%%%%%%%%%%%%%%%%%%%%%%%%%%%%%%%%%%%%%
\begin{table*}[t]
\caption{
Distribution of the electrons among the two carbon atoms $A$ and $B$ of the C$_2$ molecule in the dissociation limit. For the neutral dissociations,
only the distribution of the four $\pi$ electrons are considered since the remaining eight $\sigma$ electrons are unimportant for the study
of the dissociation. The same methods used in Fig.~\ref{fig:c2_cvb1_emin_oj_evsr} are compared. For the RHF and MCSCF wave functions, the
percentages of the distributions can be calculated analytically.
}
\begin{tabular}[b]{l|c|c|c|c}

\hline\hline
                      &     \multicolumn{3}{c|}{neutral dissociation} & ionic dissociation\\
                      & $A (\up\dn) + B (\up\dn)$   &  $A (\up\up) + B(\dn\dn)$ & $A (\dn\dn) + B(\up\up)$ & \\
\hline
RHF                   &  25 \%                    & 6.25 \%                 & 6.25 \%                & 62.5 \%\\
VMC   J$\times$SD     & $\approx$ 43 \%           & $\approx$ 13 \%         & $\approx$ 13 \%        & $\approx$ 31 \%\\
DMC   J$\times$SD     & $\approx$ 0 \%            & $\approx$ 50 \%         & $\approx$ 50 \%        & $\approx$  0 \%\\
MCSCF CAS(8,8)        & 33.33 \%                  &  33.33 \%               &  33.33 \%              &   0 \%\\
VMC J$\times$CAS(8,8) & $\approx$ 33 \%           & $\approx$ 33 \%         & $\approx$ 33 \%        & $\approx$ 0 \%\\
DMC J$\times$CAS(8,8) & $\approx$ 33 \%           & $\approx$ 33 \%         & $\approx$ 33 \%        & $\approx$ 0 \%\\

\hline\hline
\end{tabular}
\label{tab:spins}
\end{table*}
%%%%%%%%%%%%%%%%%%%%%%%%%%%%%%%%%%%%%%%%%%%%%%%%%%%%%%%%%%%%%%%%%%%%%%%%%%%%

In Fig.~\ref{fig:c2_cvb1_emin_oj_evsr}, we show the total energy curve of the all-electron C$_2$ molecule as a function of the interatomic
distance $R$ calculated in (plot a) RHF, VMC and DMC with a fully optimized Jastrow $\times$ single-determinant wave function [J$\times$SD], and
(plot b) MCSCF CAS(8,8), VMC and DMC with a fully optimized Jastrow $\times$ multi-determinant CAS(8,8) wave function [J$\times$CAS(8,8)]. In
each case, the horizontal line represents twice the energy of an isolated atom calculated with the same method, and provides a check of the
size consistency of the method. We stress that the wave functions have been optimized by energy minimization rather than variance minimization,
and in appendix~\ref{sec:sizeconsistency} we present an argument suggesting that, as regards size consistency, energy-optimized wave
functions are to be preferred over variance-optimized wave functions. For comparison, we also plot a Morse potential~\cite{Mor-PR-29},
$E_\text{Morse}(R) = E_{\exact}(R_e) + D_e (1-e^{-x^2})^2$ where $x=2\beta (R-R_e)/R_e$ and $\beta=\omega_e/(4 \sqrt{B_e D_e})$, using an
estimate of the exact energy at equilibrium $E_{\exact}(R_e)=-75.9265$ Hartree~\cite{BytRue-JCP-05} and accurate spectroscopic constants:
equilibrium distance $R_e=2.3481$~\cite{FilUmr-JCP-96}, well-depth $D_e=6.44$ eV~\cite{BytRue-JCP-05}, first vibrational frequency $\omega_e=1855$
cm$^{-1}$~\cite{NIST-BOOK-05} and rotational constant $B_e=1.81984$ cm$^{-1}$~\cite{HubHer-BOOK-79}. For analysis, we report in Table~\ref{tab:spins}
the distribution of the four $\pi$ electrons among the two carbon atoms $A$ and $B$ in the dissociation limit, the remaining eight $\sigma$
electrons being unimportant for the study of the dissociation.

We note that Sorella \textit{et al.}~\cite{SorCasRoc-JCP-07} have also reported recently QMC calculations of the potential energy curve of the C$_2$ molecule, using a pseudopotential and Jastrow antisymmetrized geminal power wave functions.

At very large interatomic distances, lack of ergodicity in the QMC calculations may be an issue, as electrons tend to remain stuck
around an atom, and nonequilibrated results can be obtained. In VMC calculations it is always possible to make large moves of the electrons between
the two atoms, as done for example in Ref.~\onlinecite{SorCasRoc-JCP-07}, but this is not possible in DMC calculations where
dynamics of the moves is specified and becomes exact only in the small time-step limit.
One can nevertheless avoid being deceived by using a large population of walkers (thereby improving the sampling of the configuration
space), looking at the evolution of the results as the bond is stretched, and performing several runs with different starting locations of
the walkers.

We first discuss the single-determinant case. RHF is of course not size consistent, and leads to a large percentage of incorrect ionic
dissociations (62.5\%). Our Jastrow factor has a multiplicatively separable form (at dissociation, it reduces to the product of the Jastrow factors
employed for the isolated atoms), so that fulfillment of size consistency in VMC calculations is only dependent on the determinantal part of the
wave function. The VMC calculation using a single determinant is not size consistent, but the Jastrow factor reduces the size-consistency error
and decreases the percentage of ionic dissociations to about 31\%. Interestingly, the DMC calculation using the nodes of a non-size-consistent
single-determinant trial wave function appears to be size consistent within the accuracy of the calculation, ionic dissociations being
absent. Moreover, examination of the distribution of electrons in the DMC calculation shows that the distribution $A(\up\dn) + B(\up\dn)$
has a vanishing probability in the dissociation limit (see Table~\ref{tab:spins}). Only the distributions $A (\up\up) + B(\dn\dn)$ and $A
(\dn\dn) + B(\up\up)$ remain at dissociation. In appendix~\ref{sec:spinsymmetry}, we show that this implies that the singlet-spin symmetry
of the ground-state is broken with an expectation value of the total spin operator $\hat{S}^2$ over the FN wave function of 2. In quantum
chemistry, it is well known that spin (and/or spatial) symmetry breaking frequently occurs in unrestricted Hartree-Fock or unrestricted
Kohn-Sham calculations at large interatomic distances where electron correlation gets stronger. Our results show that spin-symmetry breaking
can also occur in DMC calculations even using an unbroken-symmetry trial wave function, meaning that its nodal surface does not impose the
spin symmetry. One may wonder how the repeated application of a spin-independent Green function to an initial trial wave function that is a
spin eigenstate (neglecting the small spin contamination that can be introduced by imposing spin-dependent electron-electron cusp conditions
in the Jastrow factor~\cite{HuaFilUmr-JCP-98}) can result in a wave function that is not a spin eigenstate.
In fact, breaking of spin symmetry
is possible in DMC calculations because the Green function is not applied exactly but only by finite sampling. After all, the fermion-sign
problem can also been seen as resulting from the breaking of the antisymmetry of the trial wave function due to finite sampling.

We now discuss the multi-determinant case. The MCSCF calculation in a full valence CAS, i.e. CAS(8,8) for the C$_2$ molecule, is
size consistent, the corresponding atomic calculation being taken as a MCSCF CAS(4,4) calculation as in Refs.~\onlinecite{TouUmr-JCP-07}
and~\onlinecite{UmrTouFilSorHen-PRL-07}. Not surprisingly, the corresponding VMC and DMC calculations are also size consistent. The DMC energy
curve agrees closely with the reference Morse potential. For these three calculations, at the dissociation limit, the distributions $A (\up\dn)
+ B(\up\dn)$, $A (\up\up) + B(\dn\dn)$ and $A (\dn\dn) + B(\up\up)$ are obtained with equal weights, which is expected for a proper spin-singlet
wave function describing two dissociated carbon atoms, as noted in Ref.~\onlinecite{SorCasRoc-JCP-07}.

\section{Results on first-row atoms and homonuclear diatomic molecules}

%%%%%%%%%%%%%%%%%%%%%%%%%%%%%%%%%%%%%%%%%%%%%%%%%%%%%%%%%%%%%%%%%%%%%%%%%%%%%%
\begin{table*}[t]
\caption{
Total energies (in Hartree) and  well depths (in eV) of first-row atoms and homonuclear diatomic molecules at their experimental bond lengths $R_0$ (in
Bohr) using several computational methods (see text). The RHF and MCSCF calculations have been performed with the Slater CVB1 basis
set~\cite{EmaGarRamLopFerMeiPal-JCC-03}, expanding each Slater function into 14 Cartesian Gaussian functions. The QMC calculations have
been performed with the true Slater basis set rather than its Gaussian expansion. The Jastrow, CSF, orbital and exponent parameters of the
Jastrow-Slater wave functions have been optimized in VMC, and the resulting wave functions have been used in DMC.
The DMC energies are for time step $\tau=0.01$ Hartree$^{-1}$, with the exception of Ne and Ne$_2$ for which
the energies extrapolated to $\tau=0$ are given.
The well depths have been calculated consistently using single-determinant (SD) wave functions for both the molecule and the atom, or,
full valence complete active space (FVCAS) multi-determinant wave functions for both the molecule and the atom.
}
\begin{tabular}[b]{l l l l l l l l l}

\hline\hline
                    &                                          \multicolumn{8}{c}{\textbf{Atoms}}\\
                    & Li ($^2S$)  &  Be ($^1S$)  & B ($^2P$)    & C ($^3P$)   & N ($^4S$)   & O ($^3P$)   & F ($^2P$)   & Ne ($^1S$)\\
\hline
                    &                                          \multicolumn{8}{c}{\textit{Numbers of CSFs in FVCAS wave functions}}\\
                    &    1        &     2        &    2         &     2       &      1      &      1      &      1      &      1       \\
\\
                    &                                          \multicolumn{8}{c}{\textit{Total energies (Hartree)}}\\
RHF                 & -7.43271    & -14.57299    & -24.52903    & -37.68849   & -54.40060   & -74.81065   & -99.40937   & -128.54556 \\
MCSCF FVCAS         &             & -14.61663    & -24.56372    & -37.70777   &             &             &             &           \\
VMC J$\times$SD     & -7.47793(5) & -14.64972(5) & -24.62936(5) & -37.81705(6)& -54.5628(1) & -75.0352(1) & -99.7003(1) & -128.9057(1) \\
VMC J$\times$FVCAS  &             & -14.66668(5) & -24.64409(5) & -37.82607(5)&             &             &             &           \\
DMC J$\times$SD     & -7.47805(1) & -14.65717(1) & -24.63990(2) & -37.82966(4)& -54.57587(4)& -75.05187(7)& -99.71827(5)& -128.92346(3)\\
DMC J$\times$FVCAS  &             & -14.66727(1) & -24.64996(1) & -37.83620(1)&             &             &             &           \\
Estimated exact     & -7.47806$^a$& -14.66736$^a$& -24.65391$^a$& -37.8450$^a$& -54.5892$^a$& -75.0673$^a$& -99.7339$^a$& -128.9376$^a$\\
\\
             &                                          \multicolumn{8}{c}{\textbf{Molecules}}\\
             & Li$_2$ ($^1\Sigma_g^+$) & Be$_2$ ($^1\Sigma_g^+$)  & B$_2$ ($^3\Sigma_g^-$) & C$_2$ ($^1\Sigma_g^+$)  & N$_2$ ($^1\Sigma_g^+$)
             & O$_2$ ($^3\Sigma_g^-$) & F$_2$ ($^1\Sigma_g^+$) & Ne$_2$ ($^1\Sigma_g^+$) \\
\hline
                    &                                          \multicolumn{8}{c}{\textit{Interatomic distances (Bohr)}}\\

                    & 5.051$^b$   &   4.65$^c$   &   3.005$^d$  &   2.3481$^d$&    2.075$^b$ &    2.283$^b$ & 2.668$^b$  & 5.84$^g$\\
\\
                    &                                          \multicolumn{8}{c}{\textit{Numbers of CSFs in FVCAS wave functions}}\\
                    &    8        &     38       &    137       &     165       &      107   &      30      &      8      &      1       \\
\\
                    &                                        \multicolumn{8}{c}{\textit{Total energies (Hartree)}}\\
RHF                 & -14.87127   & -29.13148    & -49.08961    & -75.40154   & -108.98650   & -149.65881   & -198.76323   & -257.09105   \\
MCSCF FVCAS         & -14.89758   & -29.22111    & -49.22009    & -75.63991   & -109.13585   & -149.76453   & -198.84307   &              \\
VMC J$\times$SD     & -14.98255(5)& -29.29768(4) & -49.3457(5)  & -75.8088(5) & -109.4520(5) & -150.2248(5) & -199.4209(5) & -257.80956(2)\\
VMC J$\times$FVCAS  & -14.99229(5)& -29.33180(5) & -49.3916(2)  & -75.8862(2) & -109.4851(3) & -150.2436(2) & -199.4443(3) &              \\
DMC J$\times$SD     & -14.99167(2)& -29.31895(5) & -49.38264(9) & -75.8672(1) & -109.5039(1) & -150.2872(2) & -199.4861(2) & -257.84707(5)\\
DMC J$\times$FVCAS  & -14.99456(1)& -29.33736(2) & -49.4067(2)  & -75.9106(1) & -109.5206(1) & -150.29437(9)& -199.4970(1) &              \\
Estimated exact     & -14.995(1)  & -29.3380(4)  & -49.415(2)   & -75.9265$^f$& -109.5427$^f$& -150.3274$^f$& -199.5304$^f$& -257.8753    \\
\\
                    &                                          \multicolumn{8}{c}{\textit{Well depths (eV)}}\\
RHF                 & 0.159       &  -0.395      & 0.858        & 0.668       & 5.042           & 1.021          & -1.510       & -0.00189    \\
MCSCF FVCAS         & 0.875       &  -0.331      & 2.521        & 6.105       & 9.106           & 3.897          & 0.662        &             \\
VMC J$\times$SD     & 0.726(3)    &  -0.048(3)   & 2.367(3)     & 4.75(1)     & 8.88(1)         & 4.20(1)        & 0.55(1)      & -0.050(5)   \\
VMC J$\times$FVCAS  & 0.991(3)    &  -0.042(3)   & 2.814(6)     & 6.369(6)    & 9.78(1)         & 4.713(8)       & 1.19(1)      &             \\
DMC J$\times$SD     & 0.9679(8)   &   0.125(1)   & 2.798(3)     & 5.656(3)    & 9.583(3)        & 4.992(7)       & 1.349(6)     & 0.004(2)   \\
DMC J$\times$FVCAS  & 1.0465(6)   &   0.0767(8)  & 2.906(3)     & 6.482(3)    & 10.037(3)       & 5.187(5)       & 1.645(4)     &             \\
Estimated exact     & 1.06(4)$^c$ &   0.09(1)$^c$& 2.92(6)$^e$  & 6.44(2)$^f$ &  9.908($<$3)$^f$& 5.241($<$3)$^f$& 1.693(5)$^f$ & 0.00365$^g$\\
\hline\hline
\multicolumn{9}{l}{$^a$ Ref.~\onlinecite{ChaGwaDavParFro-PRA-93}, $^b$ Ref.~\onlinecite{HubHer-BOOK-79}, $^c$ Ref.~\onlinecite{NIST-BOOK-05}, $^d$
Ref.~\onlinecite{FilUmr-JCP-96}, $^e$ Ref.~\onlinecite{LanBau-JCP-91}, $^f$ Ref.~\onlinecite{BytRue-JCP-05}, $^g$ Ref.~\onlinecite{AziSla-CP-89}.}\\
\end{tabular}
\label{tab:energies}
\end{table*}
%%%%%%%%%%%%%%%%%%%%%%%%%%%%%%%%%%%%%%%%%%%%%%%%%%%%%%%%%%%%%%%%%%%%%%%%%%%%

\begin{figure}
\includegraphics[scale=0.35,angle=-90]{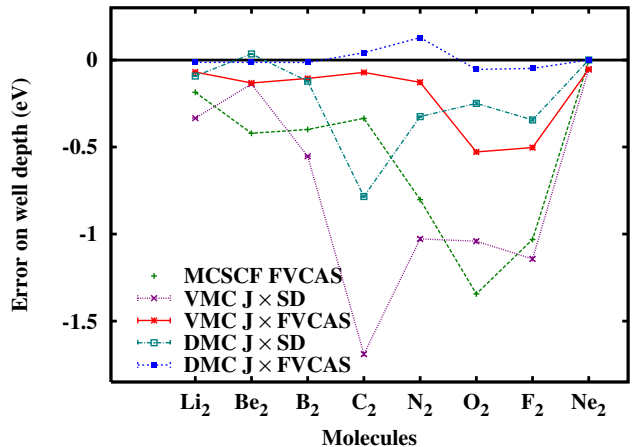}
\caption{Error in well depths of the first-row homonuclear diatomic molecules using several computational methods (see text and
Table~\ref{tab:energies}).
}
\label{fig:error_well_depth_cvb1_vmc_dmc}
\end{figure}

In Table~\ref{tab:energies}, we report total energies of the first-row atoms and homonuclear diatomic molecules at their experimental bond length
using several computational methods: RHF, MCSCF in a full valence complete active space (FVCAS), VMC with Jastrow $\times$ single determinant
[J$\times$SD] and  Jastrow $\times$ multi-determinant FVCAS [J$\times$FVCAS] wave functions (where the Jastrow, CSF, orbital and exponent
parameters have been simultaneously optimized), and DMC with the same J$\times$SD and J$\times$FVCAS wave functions. For atoms, the active
space of FVCAS wave functions consists of the $2s$ and $2p$ orbitals. For the molecules, it consists of all the orbitals coming from the $n=2$
atomic shells, i.e. $2\sigma_g 2\sigma_u 3\sigma_g 1\pi_{u,x} 1\pi_{u,y} 1\pi_{g,x} 1\pi_{g,y} 3\sigma_u$ (this is the energy ordering of the Hartree-Fock orbitals for 5 molecules out of the 8 molecules).
For the atoms Li, N, O, F and Ne, and for the dimer Ne$_2$, orbital occupations and symmetry constraints imply that the FVCAS wave functions contain only a single determinant. Thus, for these systems, the FVCAS MCSCF wave functions are identical to the RHF wave functions, and the J$\times$FVCAS wave functions are identical to J$\times$SD wave functions. The well depths (dissociation energy + zero-point energy) have been calculated consistently by using single-determinant wave functions for both the molecule and the atom, or multi-determinant FVCAS wave functions for both the molecule and the atom. The errors of the computed well depths are plotted in Fig~\ref{fig:error_well_depth_cvb1_vmc_dmc}.

The largest errors of the DMC total energy using J$\times$FVCAS wave functions are obtained for the heaviest systems and are of the order
of 15 mHartree for the atoms and 30 mHartree for the molecules. Of course, one can always improve the total energy by increasing the number of
CSFs as done for example by Brown~\textit{et al.}~\cite{BroTraLopNee-JCP-07}, but good well depths are already obtained with J$\times$FVCAS wave functions due to a compensation of errors between the atoms and the
molecule. DMC calculations using J$\times$FVCAS wave functions give well depths with near chemical accuracy (1 kcal/mol $\approx$ 0.04 eV), the largest absolute error being of about 0.1 eV for the N$_2$ molecule.
In particular, we note that, although the Be$_2$ dimer is unbound at the RHF, MCSCF and VMC level, the weak bond is well reproduced at the DMC level.
Because of the extremely weak van der Waals bond of the Ne$_2$ dimer we computed the DMC energies of Ne and Ne$_2$ at four time steps $\tau=$
0.020, 0.015, 0.010, and 0.005 Hartree$^{-1}$ and extrapolated to zero time step.
The time-step error at $\tau=0.01$ for Ne$_2$ was $-0.0023$ Hartree whereas that for Ne
was $-0.00068$ Hartree.

\section{Conclusions}
\label{sec:conclusion}

To summarize, we have extended our earlier published linear optimization method to allow for nonorthogonal orbitals. This then makes it possible to optimize all the parameters in the wave function, including the basis exponents. Moreover, by noting that the linear optimization method can be seen as an augmented Hessian method, we have shown that it is possible to minimize a linear combination of the energy and the energy variance with the linear optimization method. We have applied the method to the calculation of the full ground-state potential energy curve of the C$_2$ molecule, and we have shown that although a VMC calculation using a spin-restricted single-determinant Jastrow-Slater wave function is not size consistent, the corresponding DMC calculation using the same trial wave function is size consistent within statistical uncertainty. The price to pay for this size consistency is the breaking of the spin-singlet symmetry at dissociation: the fixed-node DMC wave function has an expectation value of $2$ for the total spin operator $\hat{S}^2$, although the spin-singlet trial wave function is an eigenstate of $\hat{S}^2$ with eigenvalue $0$. Of course, using multi-determinant FVCAS Jastrow-Slater wave functions, both the VMC and the DMC calculations are size consistent without breaking of spin symmetry. Finally, we have performed calculations on the first-row atoms and homonuclear diatomic molecules and showed that well depths can be computed with near chemical accuracy using just fully optimized multi-determinant FVCAS Jastrow-Slater wave functions.

\begin{acknowledgments}
We thank Peter Nightingale, Sandro Sorella, Andreas Savin, Frank Petruzielo, Roland Assaraf, Beno\^it Bra\"ida and Paola Gori-Giorgi for valuable discussions. We also thank Alexander Kollias, Peter Reinhardt and Roland Assaraf for providing us with the Gaussian fits of Slater basis functions of Ref.~\onlinecite{KolReiAss-JJJ-XX}.
This work was supported in part by a European Marie Curie Outgoing International Fellowship (039750-QMC-DFT), the National Science Foundation (EAR-0530301) and the DOE (DE-FG02-07ER46365). Most of the calculations were performed on the Intel cluster at the Cornell Nanoscale Facility (a member of the National Nanotechnology Infrastructure Network supported by the National Science Foundation) and at the Cornell Theory Center.
\end{acknowledgments}

\appendix

\section{A remark on size consistency and variance minimization}
\label{sec:sizeconsistency}

In this appendix, we briefly review the concept of size consistency of an electronic-structure method (see, e.g.,
Refs.~\onlinecite{DucDie-JCP-94,HelJorOls-BOOK-02,NooShaMuk-MP-05} for more details), and we give an argument for preferring energy-optimized
wave functions over variance-optimized wave functions as regards size consistency.

\subsection{Definition of size consistency}

Consider an electronic system $AB$ made of two \textit{noninteracting} fragments $A$ and $B$ (e.g., a diatomic molecule at dissociation). This
system has a Hamiltonian
\begin{equation}
\hat{H}_{AB} = \hat{H}_A + \hat{H}_B,
\end{equation}
where $\hat{H}_A$ and $\hat{H}_B$ are the Hamiltonians of the fragments, commuting with each other. If $E_A$ and $E_B$ are the (approximate)
energies of the fragments given by some method, and $E_{AB}$ is the (approximate) energy of the composite system given by the same method,
then this method is said to be \textit{size consistent} if and only if
\begin{equation}
 E_{AB} = E_A + E_B,
\end{equation}
i.e., the energy is additive.

In particular, if $\ket{A} = \hat{\psi}_A \ket{\vac}$ and $\ket{B} = \hat{\psi}_B \ket{\vac}$ are the (approximate) wave functions given by the
method under consideration where $\hat{\psi}_A$ and $\hat{\psi}_B$ are second-quantized wave operators (commuting or anticommuting with each
other), then a sufficient condition for size consistency of the method is that it leads to an (approximate) wave function for the composite
system of the product form
\begin{equation}
\ket{AB}_P = \hat{\psi}_A \hat{\psi}_B \ket{\vac},
\label{ketAB}
\end{equation}
i.e., the wave function is multiplicatively separable. However, this is not a necessary condition, as exemplified by perturbation theory (see,
e.g., the discussion in Ref.~\onlinecite{NooShaMuk-MP-05}). Also, in general, due to the nonlocality of the total spin operator $\hat{S}^2$,
one has in fact to consider a sum of products of degenerate spin-multiplet component wave functions of the fragments to accommodate non-singlet
spin symmetry, but it is sufficient to take the simple product form of Eq.~(\ref{ketAB}) for our purpose. If the wave function of the system
$AB$ has this product form, then it is easy to show that the energy variance is also additively separable
\begin{equation}
V_{AB} = V_A + V_B,
\end{equation}
where $V_{AB}=\bra{AB} (\hat{H}_{AB} - E_{AB})^2 \ket{AB}/\braket{AB}{AB}$ is the energy variance of the system $AB$, and $V_A =\bra{A} (\hat{H}_{A}
- E_{A})^2 \ket{A}/\braket{A}{A}$ and $V_B =\bra{B} (\hat{H}_{B} - E_{B})^2 \ket{B}/\braket{B}{B}$ are the energy variances of the fragments.

\subsection{Multiplicative separability of energy-optimized linear wave functions}

Before discussing variance-optimized wave functions, it is useful to repeat briefly the proof of the multiplicative separability
of energy-optimized linear wave functions, given for instance in Ref.~\onlinecite{HelJorOls-BOOK-02}.

Consider that the fragments are described by the following linearly parametrized (approximate) wave functions
\begin{subequations}
\begin{equation}
\ket{A} = \sum_i c_{iA} \ket{i_A},
\end{equation}
and
\begin{equation}
\ket{B} = \sum_j c_{jB} \ket{j_B},
\end{equation}
\label{ketAlinandketBlin}
\end{subequations}
where $\ket{i_A} = \hat{\psi}_{iA} \ket{\vac}$ and $\ket{j_B} = \hat{\psi}_{jB} \ket{\vac}$ are some many-body basis states, and the coefficients
$c_{iA}$ and $c_{jB}$ are determined by requiring the stationarity of the energy of each fragment
\begin{subequations}
\begin{equation}
\frac{\partial E_A}{\partial c_{iA}} = 2 \, \frac{\bra{i_A} \hat{H}_A - E_A \ket{A}}{\braket{A}{A}} = 0,
\end{equation}
\begin{equation}
\frac{\partial E_B}{\partial c_{jB}} = 2 \, \frac{\bra{j_B} \hat{H}_B - E_B \ket{B}}{\braket{B}{B}} = 0.
\end{equation}
\label{dEAanddEB}
\end{subequations}
Correspondingly, consider that the composite system is described by a linearly parametrized wave function in the product basis $\ket{i_A j_B}
= \hat{\psi}_{iA} \hat{\psi}_{jB} \ket{\vac}$
\begin{equation}
\ket{AB} = \sum_{ij} c_{ij} \ket{i_A j_B},
\label{ketABlin}
\end{equation}
where the coefficients $c_{ij}$ are also consistently determined by imposing the stationarity of the energy
\begin{equation}
\frac{\partial E_{AB}}{\partial c_{ij}} = 2 \, \frac{\bra{i_A j_B} \hat{H}_{AB} - E_{AB} \ket{AB}}{\braket{AB}{AB}} = 0.
\end{equation}
It is then easy to see that the product wave function $\ket{AB}_P = \hat{\psi}_A \hat{\psi}_B \ket{\vac}$ makes the corresponding energy
$E_{AB}^P$ stationary
\begin{eqnarray}
\frac{\partial E_{AB}^P}{\partial c_{ij}}  &=& 2 \, \frac{\bra{i_A} \hat{H}_{A} - E_{A} \ket{A}}{\braket{A}{A}} \,
\frac{\braket{j_B}{B}}{\braket{B}{B}}
\nonumber\\
&&+ \, 2 \, \frac{\bra{j_B} \hat{H}_{B} - E_{B} \ket{B}}{\braket{B}{B}} \, \frac{\braket{i_A}{A}}{\braket{A}{A}} = 0,
\end{eqnarray}
since both terms vanish according to Eqs.~(\ref{dEAanddEB}). The product wave function is thus a possible solution, and, if this is the actual
solution given by the method, then the method is size consistent.

As an aside, we note that the energy of the product wave function will converge exponentially to the
sum of the constituent energies as a function of the interfragment distance because the overlap
of the fragment wave functions decays exponentially.
On the other hand, the true wave function has an energy that converges only as an inverse power
to the sum of the constituent energies (Van der Waals interaction).

\subsection{Lack of multiplicative separability of variance-optimized linear wave functions}

In the case of variance-optimized linear wave functions, the coefficients $c_{iA}$ and $c_{jB}$ of the fragment wave functions of
Eq.~(\ref{ketAlinandketBlin}) are determined by requiring the stationarity of the energy variance
\begin{subequations}
\begin{equation}
\frac{\partial V_A}{\partial c_{iA}} = 2 \, \frac{\bra{i_A} ( \hat{H}_A - E_A )^2 - V_A \ket{A}}{\braket{A}{A}} = 0,
\end{equation}
\begin{equation}
\frac{\partial V_B}{\partial c_{jB}} = 2 \, \frac{\bra{j_B} ( \hat{H}_B - E_B )^2 - V_B \ket{B}}{\braket{B}{B}} = 0.
\end{equation}
\label{dVAanddVB}
\end{subequations}
Correspondingly, the coefficients $c_{ij}$ of the composite wave function of Eq.~(\ref{ketABlin}) are also determined by imposing stationarity
of the energy variance
\begin{equation}
\frac{\partial V_{AB}}{\partial c_{ij}} = 2 \, \frac{\bra{i_A j_B} ( \hat{H}_{AB} - E_{AB} )^2 - V_{AB} \ket{AB}}{\braket{AB}{AB}} = 0.
\end{equation}
In contrast to the case of energy-optimized wave functions, the product wave function $\ket{AB}_P = \hat{\psi}_A \hat{\psi}_B \ket{\vac}$ now
does not make the corresponding energy variance $V_{AB}^P$ stationary
\begin{eqnarray}
\frac{\partial V_{AB}^P}{\partial c_{ij}} &=& 2 \, \frac{\bra{i_A} ( \hat{H}_A - E_A )^2 - V_A \ket{A}}{\braket{A}{A}} \,
\frac{\braket{j_B}{B}}{\braket{B}{B}}
\nonumber\\
&& + \,  2 \, \frac{\bra{j_B} ( \hat{H}_B - E_B )^2 - V_B \ket{B}}{\braket{B}{B}} \, \frac{\braket{i_A}{A}}{\braket{A}{A}}
\nonumber\\
&& + \, 4 \, \frac{\bra{i_A} \hat{H}_{A} - E_{A} \ket{A}}{\braket{A}{A}} \,  \frac{\bra{j_B} \hat{H}_{B} - E_{B} \ket{B}}{\braket{B}{B}}
\nonumber\\
&&\not= 0,
\label{dVABP}
\end{eqnarray}
since the last term in Eq.~(\ref{dVABP}) is the product of the energy gradients of the fragments which do not now generally vanish. Thus, the
wave function minimizing the energy variance of the composite system is not a product wave function. This suggests that variance minimization
does not generally yield additively separable energies, $E_{AB} \not= E_A + E_B$.  Further, since the variance is additive for the
product wave function, if the method minimizes over a space that includes the product wave function, then
$V_{AB} \leq V_A + V_B$.
This happens by having anticorrelated energy fluctuations on the two fragments.
Of course, in the limit of exact wave functions, the gradients of the energy and of the energy variance simultaneously vanish, and
size consistency is ensured.
Consequently, the magnitude of the violation of size consistency of variance-optimized linear wave functions is expected to become
smaller as the wave function becomes more accurate.

\section{Spin-symmetry breaking in DMC for the C$_2$ molecule at dissociation}
\label{sec:spinsymmetry}

In this appendix, we show how the information on the real-space location of the electrons in DMC calculations of the C$_2$ molecule at
dissociation using a spin-restricted single-determinant trial wave function reveals that the spin-singlet symmetry of the exact ground state
is broken in the FN wave function.

For that, we need to determine the expectation value of the total spin operator $\hat{S}^2$ over the FN wave function. As we use in
the QMC calculation a real-space spin-assigned wave function, we first need to reconstitute the corresponding total wave function.
Reference~\onlinecite{HuaFilUmr-JCP-98} shows how to do so using straightforward first quantization. Here, we use the alternative
formalism of real-space second quantization.

In this formalism, the total wave function $\ket{\Gamma}$ in abstract Hilbert space corresponding to a $N$-electron real-space spin-assigned
wave function $\Psi(\b{r}_1,\b{r}_2 ,\cdots, \b{r}_N)=\braket{\b{r}_1\up,\b{r}_2\up, \cdots, \b{r}_N\dn}{\Psi}$ with $N_\up$ spin-up followed
by $N_\dn$ spin-down electrons is written as
\begin{eqnarray}
\ket{\Gamma} = \frac{1}{\sqrt{N_\up! N_\dn!}}\int d\b{r}_1 d\b{r}_2  \cdots d\b{r}_N \, \Psi(\b{r}_1,\b{r}_2 ,\cdots, \b{r}_N)
\nonumber\\
\hat{\psi}^\dag_\up(\b{r}_1) \hat{\psi}^\dag_\up(\b{r}_2) \cdots \hat{\psi}^\dag_\dn(\b{r}_N)\ket{\vac}, \,\,\,
\label{Gamma}
\end{eqnarray}
where $\hat{\psi}^\dag_\sigma(\b{r})$ is the fermionic field creation operator at point $\b{r}$ and spin $\sigma$. In Eq.~(\ref{Gamma}),
$\Psi(\b{r}_1,\b{r}_2 ,\cdots, \b{r}_N)$ is taken as antisymmetric under the exchange of two same-spin electron space coordinates and normalized
to unity, i.e. $\int d\b{r}_1 d\b{r}_2 \cdots d\b{r}_N |\Psi(\b{r}_1,\b{r}_2 ,\cdots, \b{r}_N)|^2=1$, implying that $\ket{\Gamma}$ is also
normalized to unity, $\braket{\Gamma}{\Gamma}=1$. Even if the wave function $\ket{\Psi}$ is not antisymmetric under the exchange of two
opposite-spin electrons (product of spin-up and spin-down determinants), the total wave function $\ket{\Gamma}$ is always fully antisymmetric.

To study the dissociation of the C$_2$ molecule, it is sufficient to consider only the four $\pi$ electrons. The total FN wave function
$\ket{\Gamma_\FN}$ corresponding to the spin-assigned real-space FN wave function $\Psi_\FN(\b{r}_1,\b{r}_2,\b{r}_3,\b{r}_4)$ is thus written as
\begin{eqnarray}
\ket{\Gamma_\FN} = \frac{1}{2}\int d\b{r}_1 d\b{r}_2 d\b{r}_3 d\b{r}_4 \Psi_\FN(\b{r}_1,\b{r}_2,\b{r}_3,\b{r}_4)
\nonumber\\
\hat{\psi}^\dag_\up(\b{r}_1) \hat{\psi}^\dag_\up(\b{r}_2) \hat{\psi}^\dag_\dn(\b{r}_3) \hat{\psi}^\dag_\dn(\b{r}_4)\ket{\vac},
\end{eqnarray}
with the antisymmetry constraints
\begin{subequations}
\begin{eqnarray}
\Psi_\FN(\b{r}_2,\b{r}_1,\b{r}_3,\b{r}_4)= -\Psi_\FN(\b{r}_1,\b{r}_2,\b{r}_3,\b{r}_4),
\end{eqnarray}
\begin{eqnarray}
\Psi_\FN(\b{r}_1,\b{r}_2,\b{r}_4,\b{r}_3)= -\Psi_\FN(\b{r}_1,\b{r}_2,\b{r}_3,\b{r}_4).
\end{eqnarray}
\end{subequations}

Examination of the real-space location of the electrons during DMC calculations using a spin-restricted single-determinant trial wave function
shows that, in the dissociation limit, the mixed distribution $\Psi(\b{r}_1,\b{r}_2,\b{r}_3,\b{r}_4) \Psi_\FN(\b{r}_1,\b{r}_2,\b{r}_3,\b{r}_4)$
vanishes whenever two electrons of opposite spins are in the neighborhood of the same C nucleus. In contrast, we know from VMC calculations
that the trial wave function $\Psi(\b{r}_1,\b{r}_2,\b{r}_3,\b{r}_4)$ does not forbid dissociation with two opposite-spin electrons around
the same nucleus, thus we conclude that it is the FN wave function $\Psi_\FN(\b{r}_1,\b{r}_2,\b{r}_3,\b{r}_4)$ that vanishes for this type
of dissociation. In other words, it means that at dissociation $\Psi_\FN(\b{r}_1,\b{r}_2,\b{r}_4,\b{r}_3)$ can be written as (assuming that
inversion symmetry is preserved)
\begin{eqnarray}
\Psi_\FN(\b{r}_1,\b{r}_2,\b{r}_4,\b{r}_3) &=& f_A(\b{r}_1,\b{r}_2) g_B(\b{r}_3,\b{r}_4)
\nonumber\\
&&+ g_B(\b{r}_1,\b{r}_2) f_A(\b{r}_3,\b{r}_4),
\label{PsiFNlocalized}
\end{eqnarray}
where $f_A$ and $g_B$ are antisymmetric two-electron functions localized around nuclei $A$ and $B$, respectively.

We now investigate the spin symmetry of this FN wave function. First, using the anticommutation rules of the field operators, it is
easy to show that the wave function $\ket{\Gamma_\FN}$ is an eigenstate of the spin-projection operator $\hat{S}_z = (1/2) \int d\b{r} [
\hat{\psi}^\dag_\up(\b{r}) \hat{\psi}_\up(\b{r}) - \hat{\psi}^\dag_\dn(\b{r}) \hat{\psi}_\dn(\b{r}) ]$ with eigenvalue zero
\begin{eqnarray}
\hat{S}_z \ket{\Gamma_\FN} = 0,
\end{eqnarray}
for any function $\Psi_\FN(\b{r}_1,\b{r}_2,\b{r}_4,\b{r}_3)$. The action of the total spin operator $\hat{S}^2 = \hat{S}_+ \hat{S}_- +
\hat{S}_z (\hat{S}_z-1)$ with $\hat{S}_+ = \int d\b{r} \, \hat{\psi}^\dag_\up(\b{r}) \hat{\psi}_\dn(\b{r})$ and $\hat{S}_- = \int d\b{r} \,
\hat{\psi}^\dag_\dn(\b{r}) \hat{\psi}_\up(\b{r})$ on $\ket{\Gamma_\FN}$ gives
\begin{eqnarray}
\hat{S}^2 \ket{\Gamma_\FN} = \frac{1}{2} \int d\b{r}_1 d\b{r}_2 d\b{r}_3 d\b{r}_4 [2\Psi_\FN(\b{r}_1,\b{r}_2,\b{r}_3,\b{r}_4)
\nonumber\\
 - 4\Psi_\FN(\b{r}_1,\b{r}_3,\b{r}_2,\b{r}_4)] \hat{\psi}^\dag_\up(\b{r}_1) \hat{\psi}^\dag_\up(\b{r}_2) \hat{\psi}^\dag_\dn(\b{r}_3)
 \hat{\psi}^\dag_\dn(\b{r}_4)\ket{\vac}, \,\,\,\,\,
\label{S2GammaFN}
\end{eqnarray}
where the anticommutation rules of the field operators and permutations of electron space coordinates have been used. Equation~(\ref{S2GammaFN})
shows that $\ket{\Gamma_\FN}$ is generally not an eigenstate of $\hat{S}^2$. At dissociation, it is nevertheless possible to calculate the
expectation value of $\hat{S}^2$ over $\ket{\Gamma_\FN}$
\begin{eqnarray}
\bra{\Gamma_\FN} \hat{S}^2 \ket{\Gamma_\FN} &=& 2 - \int d\b{r}_1 d\b{r}_2 d\b{r}_3 d\b{r}_4 \Psi_\FN(\b{r}_1,\b{r}_2,\b{r}_3,\b{r}_4)
\nonumber\\
&& \times \Psi_\FN(\b{r}_1,\b{r}_3,\b{r}_2,\b{r}_4)
\nonumber\\
&=& 2,
\end{eqnarray}
since the last integral vanishes due to the localized form of $\Psi_\FN(\b{r}_1,\b{r}_3,\b{r}_2,\b{r}_4)$ given in Eq.~(\ref{PsiFNlocalized}).

In conclusion, we have shown that the singlet-spin symmetry of the ground-state of the C$_2$ molecule is broken in the FN wave function at
dissociation using the nodes of a spin-restricted single-determinant wave function. The expectation value of $\hat{S}^2$ over the FN wave function
is $2$, which is identical to the value found for the lowest broken symmetry solution of the unrestricted Hartree-Fock~\cite{LepMalPel-PRA-89}
or unrestricted Kohn-Sham equations with usual approximate density functionals~\cite{GouMalSal-TCA-95}.

% BIBLIOGRAPHY---------------------------------------------
\bibliographystyle{apsrev}
\bibliography{biblio}

\end{document}